\documentclass[prb,aps,twocolumn,draft,tightenlines,%
showpacs,floatfix,superscriptaddress]{revtex4}
\usepackage{psfig}
\usepackage{amssymb}
\usepackage{amsmath}
\usepackage{colordvi}

\begin{document}

\title{Hole spin dephasing in $p$-type semiconductor quantum wells}

\author{C. L\"u}
\affiliation{Hefei National Laboratory for Physical Sciences at
  Microscale, University of Science and Technology of China, Hefei,
  Anhui, 230026, China}
\affiliation{Department of Physics,
University of Science and Technology of China, Hefei,
  Anhui, 230026, China}
\altaffiliation{Mailing Address}
\author{J. L. Cheng}
\affiliation{Hefei National Laboratory for Physical Sciences at
  Microscale, University of Science and Technology of China, Hefei,
  Anhui, 230026, China}
\author{M. W. Wu}
\thanks{Author to whom correspondence should be addressed}%
\email{mwwu@ustc.edu.cn.}
\affiliation{Hefei National Laboratory for Physical Sciences at
  Microscale, University of Science and Technology of China, Hefei,
  Anhui, 230026, China}
\affiliation{Department of Physics,
University of Science and Technology of China, Hefei,
  Anhui, 230026, China}
\altaffiliation{Mailing Address}

\date{\today}
\begin{abstract}
Hole spin dephasing time due to the D'yakonov-Perel' mechanism in
$p$-type GaAs (100) quantum wells with well separated light-hole
and heavy-hole bands
is studied by constructing and numerically solving the kinetic spin Bloch
equations. We include all the spin-conserving scattering such as the
hole-phonon and the hole-nonmagnetic impurity as well as the hole-hole
Coulomb scattering in our calculation.
Different effects such as the temperature, the hole density, the
impurity density and the Rashba coefficient
on the spin dephasing are investigated in detail.
We also show that the Coulomb scattering makes marked contribution to the
spin dephasing. The spin dephasing time can either increase or decrease with
temperature, hole/impurity density or the inclusion of
the Coulomb scattering depending on the
relative importance of the spin-orbit coupling  and the scattering.
It is also shown that due to the different spin-orbit coupling strengthes,
many spin dephasing properties of holes are quite different from those of electrons.

\end{abstract}
\pacs{72.25.Rb, 71.10.-w, 67.57.Lm, 78.90.+t}

\maketitle

\section{Introduction}

Much attention has been devoted to the spin degree of freedom
of carriers in Zinc-blende semiconductors recently due to the
possible application to the spintronic
devices.\cite{Wolf, Ziese,spintronics,das} Understanding spin
dephasing/relaxation of carriers in semiconductor quantum wells (QWs)
is one of the most important prerequisites
for the realization of such devices.   There are many studies of spin
dephasing/relaxation of electrons in $n$-type QWs where  the spin
dephasing is determined by the D'yakonov and Perel'
(DP) mechanism.\cite{dp}
 Most studies are within the framework of single-particle
approximation\cite{Optical,Lau,Song,Bronold,averkiev,Newman,kainz}
 and the spin dephasing time (SDT) can be written as\cite{Optical}
\begin{equation}
  \label{SDT_elec}
  \frac {1}{\tau} = \frac{\int^{\infty}_{0}dE_k
    (f_{k1/2}-f_{k-1/2})\tau_p(k)\overline{{\bf \Omega}^2({\bf k})} }
{2\int^{\infty}_{0}dE_k(f_{k1/2}-f_{k-1/2})}\ .
\end{equation}
Here $\tau_p(k)$ is the momentum relaxation time.
${\bf \Omega}({\bf k})$ is the DP term which is composed of
the Dresselhaus term\cite{dress} due to the
bulk inversion asymmetry (BIA) and the Rashba term\cite{rashba}
due to the structure inversion asymmetry (SIA).
 $\overline{{\bf \Omega}({\bf k})^2}$ denotes  the average of
${\bf \Omega}^2({\bf k})$ over all directions of ${\bf k}$. In GaAs QWs, the
Dresselhaus term is the leading term
and ${\bf \Omega}({\bf k})$ has the form:
\begin{eqnarray}
  \label{elec_DP}
   \Omega_x({\bf k}) &=& \gamma k_x(k_y^2-\langle k_z^2\rangle) \ ,\
   \Omega_y({\bf k}) = \gamma k_y(\langle k_z^2\rangle-k_x^2) \ ,
\nonumber\\
   \Omega_z({\bf k}) &=& 0 \ ,
\end{eqnarray}
in which $\langle k_z^2\rangle$ represents the average of the operator
$-(\partial/\partial z)^2$ over the electronic state of the lowest
 subband and is therefore $\pi^2/a^2$ under the infinite-well-depth
assumption. $\gamma$ is the Dresselhaus spin-orbit
parameter.\cite{Optical,aronov}
It is noted that Eq.\ (\ref{SDT_elec}) is
valid only when ${\bf{|\Omega |}} \tau_p \ll 1$ and the
scattering is elastic. In this limiting
case, the angular rotation of electron spin over time $\tau_p$ is
small and spin relaxation occurs as a result of a number of accidental
small rotations.\cite{Optical} This approach captures the lowest (first)
order of the anisotropy due to the fact that ${\bf \Omega}(-{\bf
k})=-{\bf \Omega}({\bf k})$.

It is shown recently by Wu {\em et al.} from a full many-body
microscopic
approach\cite{wu1,wu2,wu3,Weng,MQ,strain,wu4,transport} that the
single-particle approach is inadequate in accounting for the spin
dephasing/relaxation. The momentum dependence of the effective
magnetic field (the DP term), and even the momentum dependence of
the spin diffusion rate along the spacial gradient\cite{wu4} or
the random spin-orbit coupling,\cite{sherman}
serve as inhomogeneous broadenings.\cite{wu2,wu3} In the presence
of the inhomogeneous broadening, any scattering, including the
carrier-carrier Coulomb scattering (beyond the Hartree-Fock
self-energy from the Coulomb interaction), can cause irreversible
dephasing. This many-body approach  takes account of the
inhomogeneous broadening not only from different directions of
${\bf \Omega}({\bf k})$ (not only $-|{\bf\Omega}({\bf k})|$ and
$+|{\bf\Omega}({\bf k})|$), but also from the modulus of the DP
effective field, {\em i.e.}, $|{\bf \Omega}({\bf
k})|$.\cite{comment} Moreover, this approach also takes full
account of the counter effect of the scattering to the
inhomogeneous broadening instead of only the lowest-order elastic
scattering: The scattering tends to drive carriers to more
homogeneous states and therefore suppresses the inhomogeneous
broadening induced by the DP term. Finally, this approach is valid
even when $|{\bf \Omega}({\bf k})|\tau_p\gtrapprox 1$ and is applicable
to systems far away from equilibrium ({\em eg.}, systems with high
spin polarization\cite{Weng} and/or in the presence of high electric
field parallel to QWs).\cite{MQ,strain} Using this method, Weng
and Wu performed a systemic studies of spin dephasing  in $n$-type
GaAs (100) QWs at high temperatures and showed that the effects
beyond the single-particle approach Eq.\ (\ref{SDT_elec}) are
dominant even for systems near equilibrium.\cite{Weng} These
effects include the many-body effects, the inhomogeneous
broadening induced spin dephasing and the counter effect of the
scattering to the inhomogeneous broadening.
For small well width, the calculated electron SDTs  using this
microscopic many-body theory increase with temperature and are in
agreement with the experiment both qualitatively and quantitatively,
while the SDTs of earlier simplified treatment drop dramatically with
temperature and are one order of magnitude larger than the experiment
data.\cite{Weng} For larger well width, the SDT may first increase
then decrease with temperature.\cite{strain} These properties
come from the competing effects between the DP term and the
scattering.

Although there are extensive investigations on the spin relaxation/dephasing
of electrons, investigations on the spin relaxation/dephasing of holes
in $p$-type semiconductor QWs are relatively
limited.\cite{sham,bastard} Nevertheless, knowledge of the spin
relaxation/dephasing of holes in $p$-type QWs is very important
to the assessment of the feasibility of hole-based spintronic
devices. This is because  a possible way to achieve high
electronic spin injection
without the conductance mismatch\cite{mismatch}
is to use magnetic semiconductors as spin source and
most magnetic semiconductors are $p$-type at high temperature.\cite{ohno}
Very recently there are some reports
on the hole spin relaxation/dephasing.\cite{tang,glazov,kim,Yu} All the
theoretical calculations in these
works are within the framework of the single-particle approximation
Eq.\ (\ref{SDT_elec}).\cite{kim,Yu}

It has been shown in electron systems that Eq.\ (\ref{SDT_elec})
is inadequate in accounting for the spin dephasing.
Moreover, the electronic states and spin-orbit coupling of holes
are very different from those of electrons.\cite{bir,Winkler_book,Rashba_coef1}
In bulk material, the $\Gamma$-point degeneracy of the heave hole (HH) and
the light hole (LH) makes the hole spin relaxation in the same order of
the momentum  relaxation (100\ fs).\cite{tang} This degeneracy is lifted
in QWs.  Under the parabolic approximation,  the HH and LH bands can be
treated independently for QWs of small well width. Unlike the
conduction band where the
DP term mainly comes from the BIA contribution in GaAs QWs,
in $p$-type GaAs QWs, the SIA contribution is usually the dominant
one. It is noted that in hole system the relation $|{\bf \Omega}|
\tau_p\ll 1$ is usually {\em unsatisfied} due to the strong
spin-orbit coupling  and consequently the validity of
Eq.\ (\ref{SDT_elec}) is even more questionable. Therefore, in this
paper we investigate the hole spin dephasing using our full many-body
microscopic approach. We calculate the SDT of the HH and LH
by numerically solving the many-body spin kinetic Bloch
equations with all the scattering explicitly included. Then we discuss how the
temperature, the hole density, the Coulomb scattering,
the Rashba coefficient and the impurity
density affect the SDT.
We show that the eariler treatment based on the single particle
approximation is not valid in hole systems and unlike the case of electrons
where the scattering ``always'' raises the SDT at low-spin polarization,
the scattering can either enhance or suppress the SDT
of holes based on the relative importance of the Rashba term and the scattering.

The paper is organized as follows: In Sec.\ II we set up our model and
kinetic equations.  Then in Sec.\ III we present our numerical
results. We first show the time evolution of the spin signal in
Sec.\ III A.
In Sec.\ III B we investigate how the temperature
affects the spin dephasing. The Coulomb scattering,
the impurity density and hole density dependence of the SDT are
discussed separately in Sec.\ III B,  C and  D. We
conclude  in  Sec.\ IV. In Appendix A we show the effect of the
scattering to spin dephasing when it is much weaker than the
spin-orbit coupling strength. In Appendix B we present a simplified
analytical analysis of the SDT and show the different effects of
the scattering to the spin dephasing at strong/weak scattering regime.

\section{Kinetic Spin Bloch Equations}

We start our investigation from a $p$-doped (100) GaAs QW of well width $a$.
The growth direction is assumed to be along the $z$ axis. A
moderate magnetic field ${\bf B}$ is applied along the $x$ axis
 (in the Voigt configuration). Here
we assume only the lowest subband is populated. It is noted that for
two-dimensional hole system, the lowest subband is HH-liked.
By applying a suitable strain, the lowest subband can be LH-liked.
We assume the confinement is large enough so that the
HH and LH bands are well separated and we may consider the HHs and LHs
separately.
With the DP term included, the Hamiltonian of the holes
can be written as:
\begin{eqnarray}
H_\lambda& =&\sum_{{\bf k}\sigma\sigma^\prime}\{\varepsilon_{{\bf
      k}\lambda}\delta_{\sigma\sigma^\prime}
+[g_{\lambda}\mu_B{\bf B} + {\bf \Omega}^\lambda
({\bf k})]\cdot\frac{\mbox{\boldmath$\sigma$\unboldmath}
_{\sigma\sigma^\prime}}{2} \} c^{\dagger}_{\lambda{\bf
      k}\sigma}c_{\lambda{\bf k}\sigma^{'}}\nonumber\\
&&\mbox{}+H_I \ .
  \label{total_Hamiltonian}
\end{eqnarray}
Here $\lambda = LH, HH$ denotes the LH  or HH
state, $\sigma = +, -$ stands for the spin. $\varepsilon_{{\bf k}\lambda} =
{\bf k}^2/2m^\ast_{\lambda}$ is the
energy of hole with wave vector ${\bf k}$ and effective mass
$m^\ast_{\lambda}$. \boldmath$\sigma$\unboldmath\  are the Pauli matrices. The
DP term is mainly from the Rashba term. For
(100) GaAs QWs, we have
\begin{eqnarray}
  \label{HH_DPx}
\Omega^{HH}_x({\bf k})&=&2{ E}_z[\gamma_{53}^{7h7h}k_{\|}^2k_y
+\gamma_{54}^{7h7h}k_y(k_y^2-3k_x^2)] \ , \\
   \Omega^{HH}_y({\bf k})&=& -2{ E}_z[\gamma_{53}^{7h7h}k_{\|}^2k_x
+\gamma_{54}^{7h7h}k_x(k_x^2-3k_y^2)] \ , \\
   \Omega^{HH}_z({\bf k}) &=& 0 \ ,
  \label{HH_DPz}
\end{eqnarray}
for HHs and
\begin{eqnarray}
  \label{LH_DPx}
   \Omega^{LH}_x({\bf k}) &=& 2{ E}_z[\gamma_{52}^{6l6l}
\langle k_z^2\rangle k_y +
   \gamma_{53}^{6l6l}k_{\|}^2k_y\nonumber\\
&&\hspace{0.6cm}
  +\gamma_{54}^{6l6l}k_y(k_y^2-3k_x^2)] \ ,\\
\Omega^{LH}_y({\bf k})& =& -2{ E}_z[\gamma_{52}^{6l6l}
\langle k_z^2\rangle k_x
   + \gamma_{53}^{6l6l}k_{\|}^2k_x
   \nonumber\\
\label{LH_DPy}
   &&\hspace{0.6cm}+\gamma_{54}^{6l6l}k_x(k_x^2-3k_y^2)] \ ,\\
\Omega^{LH}_z({\bf k}) &=& 0
  \label{LH_DPz}
\end{eqnarray}
for LHs.\cite{Winkler_book} It is seen from these equations that
the magnitude of the  Rashba term can
be tuned by means of an external gate voltage which changes the electric field
${ E}_z$ in the sample.\cite{Wollrab,Luo,Nitta,Heida}
$\gamma_{53}^{7h7h}$, $\gamma_{54}^{7h7h}$,
$\gamma_{52}^{6l6l}$, $\gamma_{53}^{6l6l}$ and $\gamma_{54}^{6l6l}$
in Eqs.\ (\ref{HH_DPx}-\ref{LH_DPz}) are the
Rashba coefficients:\cite{Winkler_book}
$\gamma_{53}^{7h7h} =
\frac{3}{4}\frac{e\hbar^4}{m_0^2}\gamma_3(\gamma_2-\gamma_3)
(\frac{1}{\Delta_{hl}^2} -\frac{1}{\Delta_{hs}^2})$,
$\gamma_{54}^{7h7h} =
\frac{3}{4}\frac{e\hbar^4}{m_0^2}\gamma_3(\gamma_2+\gamma_3)
(\frac{1}{\Delta_{hl}^2} -\frac{1}{\Delta_{hs}^2})$,
$\gamma_{52}^{6l6l} =
-3 \frac{e\hbar^4}{m_0^2}\frac{\gamma_2\gamma_3}
{\Delta_{ls}^2}$,
$\gamma_{53}^{6l6l}=
\frac{3}{2}\frac{e\hbar^4}{m_0^2}\gamma_3
[(\frac{1}{2\Delta_{hl}^2} +\frac{1}{\Delta_{ls}^2})\gamma_2 +
\frac{\gamma_3}{2 \Delta_{hl}^2}]$ and
$\gamma_{54}^{6l6l} =
-\frac{3}{4}\frac{e\hbar^4}{m_0^2}\frac{\gamma_3(\gamma_2-\gamma_3)}
{\Delta_{hl}^2}$
in which  $\Delta_{hl}$, $\Delta_{hs}$ and $\Delta_{ls}$
present the energy gaps between the HH and the LH bands,
the HH and the split-off  bands, the LH and
the split-off  bands respectively:
\begin{eqnarray}
\label{delta_hl}
\Delta_{hl}& =& 4\gamma_2 \frac{\hbar^2 \langle
  k_z^2\rangle }{2m_0}\ ,\\
\label{delta_hs}
\Delta_{hs}& =&\Delta_0 -(\gamma_1-2\gamma_2)
\frac{\hbar^2  \langle   k_z^2\rangle }{2 m_0}\ ,\\
\label{delta_ls}
\Delta_{ls}& =& \Delta_0 -(\gamma_1+2\gamma_2) \frac{\hbar^2\langle
  k_z^2\rangle }{2 m_0}\ ,
\end{eqnarray}
with $\Delta_0$ representing the energy gap of the split-off band (from the
$\Gamma$-point of the valence band). $\gamma_1$, $\gamma_2$
and  $\gamma_3$ are the  Luttinger parameters.
From Eqs.\ (\ref{HH_DPx}-\ref{LH_DPz}) one can see that the HH
Rashba terms include only the cubic terms whereas the LH ones include
both the cubic and the linear terms. The ratio of the cubic and
the linear terms  depends on the well width $a$:  $\langle
k_z^2\rangle$ in the linear  terms
decreases with $a^2$. Futhermore, one can see from Eqs.\
(\ref{delta_hl}) and (\ref{delta_ls})
 that ${\Delta_{hl}}$ decreases faster with $a$ than
${\Delta_{ls}}$, which makes $\gamma_{53}^{6l6l}$ and
$\gamma_{54}^{6l6l}$ increase faster with $a$ than
$\gamma_{52}^{6l6l}$. Therefore, the cubic terms weighted by
$\gamma_{53}^{6l6l}$ and $\gamma_{54}^{6l6l}$ increase faster
with well width  than the linear terms weighted by $\gamma_{52}^{6l6l}$.
In brief, when $a$ is small, both the linear
and the cubic terms are important; When $a$ gets larger, the cubic
terms are the leading terms.

The interaction Hamiltonian $H_I$ in Eq.\ (\ref{total_Hamiltonian})
 is composed of the hole-hole Coulomb
interaction  and hole-phonon scatering, as well
as hole-impurity scattering. Their expressions can be found
in textbooks.\cite{Haug,Ivchenko}

We construct the many-body kinetic spin Bloch equations by the
non-equilibrium Green function method as follows:
\begin{equation}
  \label{Bloch_eq}
  \dot{\bf \rho}_{{\bf k}\lambda,\sigma\sigma^\prime} = \dot{\bf \rho}_{{\bf
          k}\lambda,\sigma\sigma^\prime}|_{coh} +
\dot{\bf \rho}_{{\bf k}\lambda,\sigma\sigma^\prime}|_{scatt}
\end{equation}
with ${\bf \rho}_{{\bf k}\lambda,\sigma\sigma^\prime}$ representing
the single-particle density matrix elements. The diagonal elements
${\bf \rho}_{{\bf k}\lambda,\sigma\sigma} \equiv {f}_{{\bf
    k}\lambda,\sigma}$ describe the hole distribution functions of
wavevector ${\bf k}$ , state $\lambda$ and spin $\sigma$. The
off-diagonal elements ${\bf \rho}_{{\bf k},\lambda,+ -}
={\bf \rho}^\ast_{{\bf k},\lambda,- +}\equiv
{\bf \rho}_{{\bf k}\lambda} $ describe the inter-spin-band
correlations for the spin coherence. $\dot{\bf \rho}_{{\bf
    k}\lambda,\sigma\sigma^\prime}|_{coh}$ describe the coherent
spin precessions around the applied magnetic field ${\bf B}$ in the
Voigt configuration, the
effective magnetic field ${\bf \Omega}^\lambda({\bf k})$
 as well as the effective
magnetic field from the hole-hole Coulomb interaction in the
Hartree-Fock approximation and can be written as:
\begin{widetext}
\begin{eqnarray}
  \label{coh_f}
\left.\frac{\partial f_{{\bf k}\lambda,\sigma}}{\partial t}\right|_{coh}
& =& - 2 \sigma\{[g_\lambda\mu_BB + \Omega_x^\lambda({\bf k})]\mbox{Im}
\rho_{{\bf k}\lambda}+\Omega_y^\lambda({\bf
      k})\mbox{Re}\rho_{{\bf k}\lambda}\}
+ 4 \sigma \mbox{Im} \sum\limits_{\bf q} V_{\bf q}
  \rho_{{\bf k}+{\bf q}\lambda}^\ast \rho_{{\bf k}\lambda} \ ,\\
\left.\frac{\partial \rho_{{\bf k}\lambda}}{\partial t}\right|_{coh}  &=&
  \frac{1}{2} [ig_\lambda\mu_BB + i\Omega_x^\lambda({\bf k})+\Omega_y
^\lambda({\bf k})](f_{{\bf k}\lambda, +} - f_{{\bf k}\lambda, -})
+ i \sum\limits_{\bf q} V_{\bf q} [(f_{{\bf k}+{\bf q} \lambda, +} -
  f_{{\bf k}\lambda, -}) \rho_{{\bf k}\lambda}\nonumber\\
&&\mbox{}- \rho_{{\bf k}+{\bf q}\lambda}(f_{{\bf k}\lambda,+}- f_{{\bf k}
\lambda, -})]\ .
\label{coh_rho}
\end{eqnarray}
\end{widetext}
$\dot{\bf \rho}_{{\bf
    k}\lambda,\sigma\sigma^\prime}|_{scatt}$ in Eq.\ (\ref{Bloch_eq})
 denote the hole-hole Coulomb, hole-phonon and hole-impurity
scattering. The expressions of these scattering terms and the details
of solving these many-body kinetic spin Bloch equations can be found in
Ref. [\onlinecite{MQ}].

\section {Numerical Results}

We numerically solve the kinetic spin Bloch equations and obtain temporal
evolution of the hole distribution $f_{{\bf k}\lambda,\sigma}(t)$ and the spin
coherence $\rho_{{\bf k}\lambda}(t)$.
We include the hole-phonon and the hole-hole
scattering throughout our computation. As we concentrate on the
relatively high-temperature regime ($T\ge 120$\ K), we only
include the hole-LO-phonon scattering.
The hole-impurity scattering is included when stated. As
discussed in the previous papers,\cite{wu1,wu2,Kuhn,  Haug}
the irreversible spin dephasing can be well defined by the slope of
the envelope of the incoherently summed spin coherence
\begin{equation}
  \label{sum_rho}
  \rho_\lambda= \sum_{\bf k}|\rho_{{\bf k}\lambda}(t)| \ .
\end{equation}
The material parameters of GaAs in our calculation are tabulated in Table\
\ref{parameter} where $\Omega_{\mbox{\tiny LO}}$ represents the
LO phonon frequency and $\kappa_\infty$ ($\kappa_0$) is
the optical (static) dielectric constant.\cite{LB,Winkler_book}
 Our main results are plotted in
Figs.\ 1-6. In these calculations the width of the QW is
chosen to be 5\ nm unless otherwise specified; the initial spin
polarization $P_\lambda=(N_{\lambda,+}-N_{\lambda,-})
/(N_{\lambda,+}+N_{\lambda,-})$ is 2.5\ \% with
\begin{equation}
  \label{sum_N}
  N_{\lambda,\sigma}=\sum_{\bf k}f_{{\bf k}\lambda,\sigma} \ ,
\end{equation}
representing the hole density of $\sigma$-spin band; the
magnetic field $B=4$\ T, and the Rashba coefficient
$\gamma_{54}^{7h7h}E_zm_0$ is taken to $0.5$\ nm when $a=5$\ nm.

\subsection{Temporal evolution of the spin signal}

We first study the temporal evolution of the spin signal in a GaAs
QW at $T = 300$\ K. In Fig.\ 1 we show the typical evolution of
the HH densities in the spin-up and -down bands together with the
incoherently summed spin coherence for the total HH density
$N_{HH,+}+N_{HH,-}=4 \times 10^{11}$\ cm$^{-2}$ and impurity
density $N_i = 0$. It is seen from the figure that the hole
densities  in the spin-up and -down bands and the incoherently
summed spin coherence oscillate due to the presence of the
magnetic field. From the slope of the envelope of the incoherently
summed spin coherence, one is able to deduce the SDT.

\begin{figure}
  \centerline{
  \psfig{figure=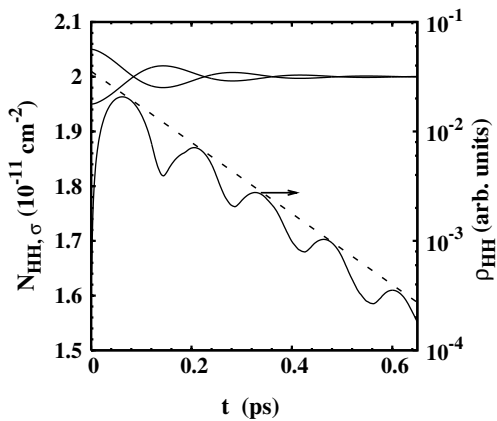,height=5.5cm,angle=0}}
  \caption{Typical hole densities of spin-up and spin-down bands and the
    incoherently summed spin coherence $\rho_{HH}$ {\em vs.} time $t$
shown for the case of the HHs. Note that
    the scale of the spin coherence is on
    the right side of the figure. The dotted line represents the
slope of the envelope of $\rho_{HH}$.}
\end{figure}

\begin{table}[htbp]
\caption{Parameters used in the calculation.}
\begin{tabular}{lllllllllllllll}\hline\hline
$\kappa_{\infty}$&\mbox{}&$10.8$ &\mbox{}&\mbox{}
&\mbox{}&$\kappa_0$&\mbox{}&12.9&\mbox{}&\mbox{}&\mbox{}&
$\gamma_1$&\mbox{}&6.85\\
$\Omega_{\mbox{\tiny LO}}$&\mbox{}&$35.4$\ meV
&\mbox{}&\mbox{}&\mbox{}&$m_{LH}$&\mbox{}
&$0.067 m_0$&\mbox{}&\mbox{}&\mbox{}&$\gamma_2$&\mbox{}&2.1\\
$\Delta_0$&\mbox{}&$0.341$\
eV&\mbox{}&\mbox{}&\mbox{}&$m_{HH}$&\mbox{}
&$0.5 m_0$&\mbox{}&\mbox{}&\mbox{}&$\gamma_3$&\mbox{}&2.9\\
$E_{g}$&\mbox{}&$1.55$\
eV&\mbox{}&\mbox{}&\mbox{}&$g_{LH}$&\mbox{}
&$1.2$&\mbox{}&\mbox{}&\mbox{}&$g_{HH}$&\mbox{}&3.6\\
\hline\hline
\end{tabular}
\label{parameter}
\end{table}

\subsection{Temperature dependence of the SDT}

We now turn to study the temperature dependence of the SDT at
different impurity densities $N_i$. We plot the SDTs of the LH and
the HH in Fig.\ 2(a) and (b) as functions of temperature. The
total LH and HH densities $N_{LH}$ and $N_{HH}$  are taken to be
$N_h=4 \times 10^{11}$\ cm$^{-2}$. One finds from Fig.\ 2(a) that
for the LH, when there are no impurity $N_i=0$ or low impurities
$N_i=0.1N_h$, the SDT first decreases then increases with
temperature. The minimum occurs at smaller temperature for higher
impurity densities:  140\ K when $N_i=0.1N_{h}$ and 200\ K when
$N_i=0$. When the impurity density $N_i = N_h$, the SDT increases
with temperature monotonically. These temperature dependences are
quite different from those of electrons in QWs with the same
electron density  and initial spin polarization where the SDT {\em
increases} monotonically with temperature.\cite{Weng}

It is noted that the property of spin dephasing is quite different when
 $|{\bf\Omega}| \tau_p \gtrapprox 1$ and  $|{\bf\Omega}| \tau_p \ll 1$.
When $|{\bf\Omega}| \tau_p \gtrapprox 1$, the
scattering is weak in comparison to the DP effective field
(inhomogeneous broadening) and the counter effect of the
scattering to the inhomogeneous broadening is unimportant or can be ignored. In the
presence of inhomogeneous broadening, adding a new scattering
provides an additional dephasing channel.\cite{allen,wu2,Haug}
This effect has been revealed in detail in Appendix A.  Therefore,
the scattering in this regime provides a  spin dephasing channel
and the increase of the temperature leads to a stronger scattering
and consequently a faster spin dephasing. Moreover,  the increase
of the temperature drives holes to a higher ${\bf k}$-state, and
holes experience a larger $|{\bf \Omega}({\bf k})|$, {\em i.e.}, a
stronger inhomogeneous broadening. This tends to reduce the SDT
too.  Therefore, the SDT decreases with temperature when $|{\bf
\Omega}| \tau_p \gtrapprox 1$. When ${\bf{|\Omega |}} \tau_p \ll 1$, the
scattering is strong in comparison to the DP term. Hence the
counter effect of the scattering to the inhomogeneous broadening
cannot be ignored any more. As the scattering tends to drive carriers to
more homogeneous states in ${\bf k}$-space, it tends to increase
the SDT. Therefore, whether the SDT  increases or decreases with
temperature depends on the competition between the scattering and
the DP term. It  will be shown later that when the linear part in
the DP term is dominant, the increase of the inhomogeneous
broadening with temperature is relatively slower than that of the
scattering and the SDT  increases with temperature. Nonetheless,
when the cubic part in the DP term is dominant, the increase of
the inhomogeneous broadening with temperature turns out to be
faster than the increase of the scattering and the SDT decreases
with temperature.

For electrons in GaAs QWs, the spin-orbit
coupling is not very strong. $|{\bf \Omega}|
\tau_p$ is usually much smaller than 1 (typically $|{\bf \Omega}|
\tau_p = 0.016$ at $T = 100$\ K, $a=15$\ nm, and the total electron density
$N_{e,+}+N_{e,-} = 4 \times 10^{11}$\ cm$^{-2}$).
Therefore when the linear (cubic) term in Eq.\ (\ref{elec_DP}) is dominant,
the SDT of electrons increases (decreases) with temperature.\cite{Weng, strain}

\begin{table}[htbp]
\caption{Rashba coefficients [unit: nm$/(E_zm_0)$].}
\begin{tabular}{lllllllllllllll}\hline\hline
\ \ \ \ \ \ \ &\mbox{}& $\gamma_{52}^{6l6l}$ &\mbox{} &$\gamma_{53}^{6l6l} $ &\mbox{}
&$\gamma_{54}^{6l6l} $ &\mbox{} & $\gamma_{53}^{7h7h} $ &\mbox{} &
$\gamma_{54}^{7h7h} $ &\mbox{} & \\\hline
$a= 5$\ nm &\mbox{}& $-0.193$\ &\mbox{}& 0.650\ &\mbox{}& 0.089\
&\mbox{}& $-0.080$\ &\mbox{}&  0.500\ &\\
$a = 7$\ nm&\mbox{}&$-0.156$\ &\mbox{}&2.21\ &\mbox{}&0.341\ &
\mbox{}&$-0.330$\ & \mbox{}&2.07\ &\\
\hline\hline
\end{tabular}
\label{parameter_Rashba}
\end{table}

\begin{figure}
  \centerline{
  \psfig{figure=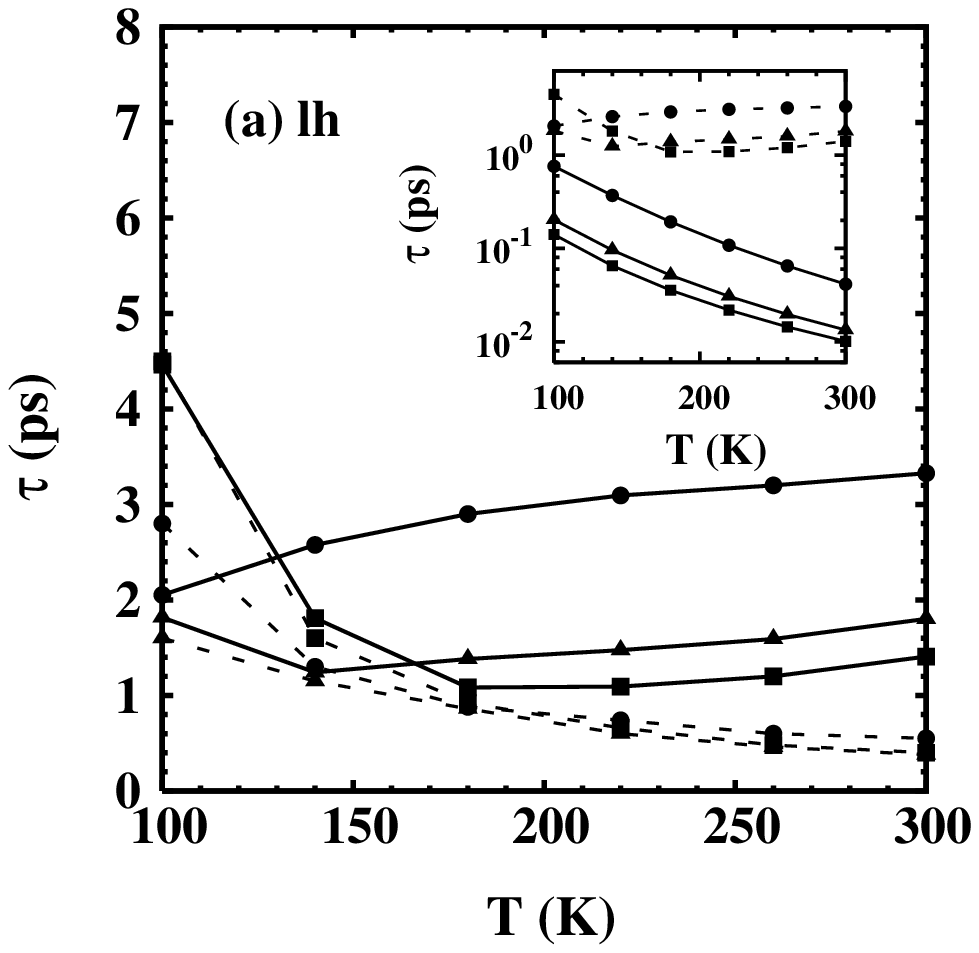,height=5.5cm,angle=0}}
  \vskip 0.5pc
\centerline{\psfig{figure=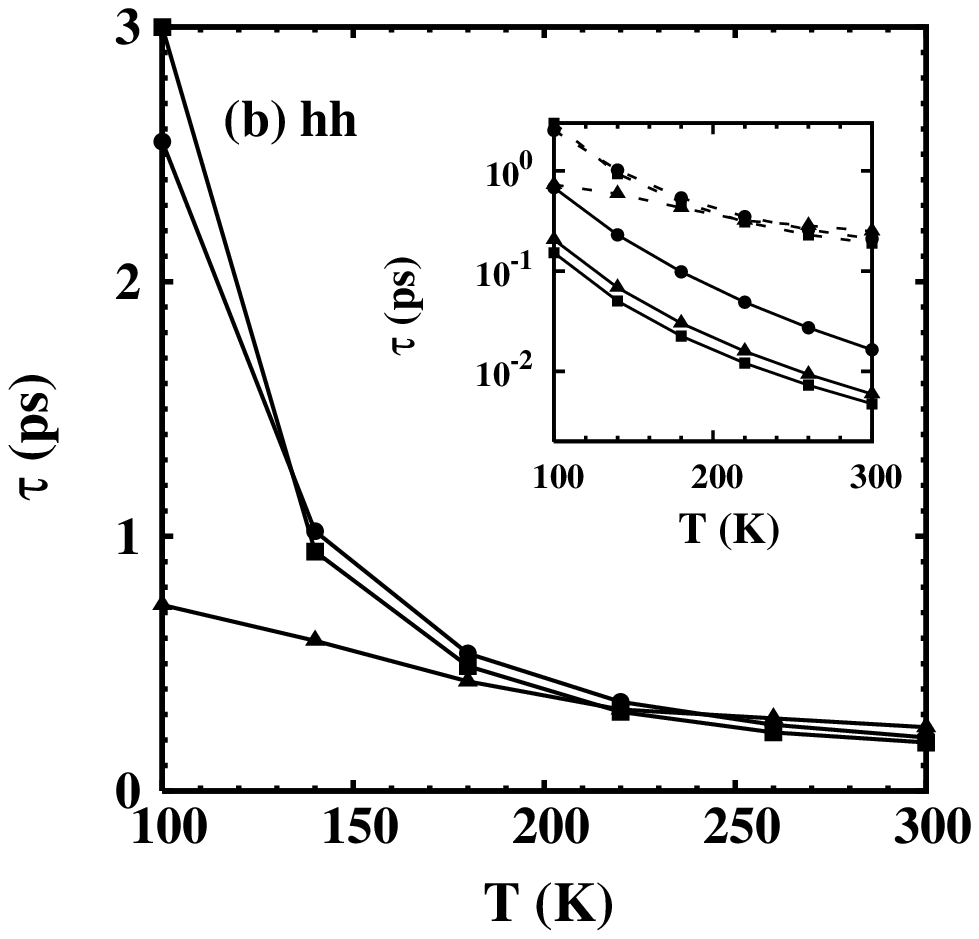,height=5.5cm,angle=0}}
  \caption{SDT $\tau$  of LHs [Fig.\ 2(a)] and
  HHs [Fig.\ 2(b)] {\em vs.} temperature $T$ at
different impurity densities. {\tiny $\blacksquare$}: $N_i =
0$; {$\blacktriangle$}: $N_i= 0.1N_h$; and
{$\bullet$}: $N_i = 1.0N_h$. The dashed curves in
    Fig.\ 2(a) are the corresponding curves but at bigger well width
$a = 7$\ nm. The SDTs calculated from
the simplified treatment (solid curves) and our many-body treatment (dashed
curves) are plotted in the insets for comparison.}
\end{figure}

Situations are more complicated for hole system due to the strong
spin-orbit coupling. The Rashba coefficients of the coupling
in Eqs.\ (\ref{HH_DPx}-\ref{LH_DPz})
are listed in Table\ \ref{parameter_Rashba}. For LHs, when $a=5$\ nm,
${\gamma_{52}^{6l6l}E_z}{m_0}\frac{\langle
  k_z^2\rangle}{\langle k_\|^2 \rangle}$ changes from $ -3.36$\ nm to
$-1.46$\ nm when the temperature changes from 100\ K to 300\ K. Here
$\langle k_\|^2 \rangle$ represents the average of $k_\|^2$. It is seen
from Table\ \ref{parameter_Rashba} that both
$\gamma_{53}^{6l6l}$ and
$\gamma_{54}^{6l6l}$ are smaller than
$-\gamma_{52}^{6l6l}\frac{\langle
  k_z^2\rangle}{\langle k_\|^2 \rangle}$. Therefore the linear terms
in Eqs.\ (\ref{LH_DPx}) and (\ref{LH_DPy}) are dominant.
From these coefficients, one may further find that when there is no
impurity, the spin-orbit coupling for LHs is one or two orders of
magnitude larger than that of electrons. Consequently
neither $|{\bf \Omega}^{LH}| \tau_p \gtrapprox 1$ nor  $|{\bf\Omega}^{LH}|
\tau_p \ll 1$ is satisfied here.
The value of $|{\bf\Omega}^{LH}|
 \tau_p$ is usually slightly smaller than 1. In this regime,
both the competing effects of the scattering addressed above can not be neglected.
Therefore the temperature dependence of the SDT depends on the competition
between the effect of the increase of the spin dephasing due to the
increase of the inhomogeneous broadening and the increase of the
scattering with temperature (The latter provides additional spin dephasing channel)
(Effect I)
and the effect of the decrease of the spin dephasing due to
the counter effect from the increase
of scattering which suppresses the inhomogeneous broadening (Effect II).
The results for impurity free case shown in Fig.\ 2(a) indicate that when
$T < 200$\ K and the total scattering is not so strong, Effect I
 is more important and hence the SDT
decreases with $T$. When temperature keeps increasing and the
total scattering is further enhanced, the counter effect of the
scattering to the inhomogeneous broadening (Effect II) becomes
more important and the SDT increases with $T$. Comparing with our
previous works,\cite{Weng, strain} one further finds that the
absolute value of the SDT of LHs is one or two orders of magnitude
smaller than the SDT of electrons. This can be easily understood
from the fact that the Rashba coefficients here are larger.

Now we include the hole-impurity scattering
with the impurity density $N_i = 0.1 N_{h}$. As expected, when the
total scattering becomes stronger, the counter effect of the scattering to the
inhomogeneous broadening takes the leading place easier and the
SDT starts to increase with temperature earlier than the impurity-free case.
When $N_i = 1.0 N_{h}$, the total scattering is further
enhanced. Now if one uses the hole-impurity scattering and the hole-phonon
scattering to calculate the momentum relaxation time, and takes the
lowest order of $\tau_p^{-1}({\bf k})$ and ${\bf \Omega}^{LH}
({\bf k})$ after expanding them
over the function $A_l(\theta_{\bf k})$ defined in Eq.\
(\ref{expand_rho}),  one gets the typical value of $|{\bf
  \Omega}^{LH}({\bf k})| \tau_p({\bf k})$ at the average of $k$ to be
$0.11$ at $T = 100$\ K. It has already
  entered the regime of strong scattering, and
similar to the case of electrons when the linear part of the DP term
is dominant, the SDT increases monotonically with temperature.

When the well width becomes larger, the cubic term becomes more
important. For example, when $a= 7$\ nm,
${\gamma_{52}^{6l6l}E_z}{m_0}\frac{\langle
k_z^2\rangle}{\langle k_\|^2 \rangle}$ changes from $ -1.39$\ nm  to
$-0.60$\ nm when the temperature changes from 100\ K to 300\ K.
One can see from Table\ \ref{parameter_Rashba} that the cubic terms
weighted by $\gamma_{53}^{6l6l}$ are dominant.
The SDT in this case is plotted in Fig.\ 2(a) as dashed
curves for comparison. One can see that
now the SDT  decreases monotonically with temperature.
It is because the increase of the inhomogeneous broadening with
temperature is much faster when the cubic term in the DP term is
dominant. Therefore Effect I always surpasses Effect II with the
increase of temperature and
the SDT decreases monotonically with $T$.
The same situation happens in the case of HHs where there is only
cubic term in the DP term. It is seen from Fig.\ 2(b)
that the SDT  decreases monotonically with temperature even when
$a=5$\ nm. This is consistent to the electron case when the cubic
term is dominant or the only term (bulk case)
where the SDT also decreases monotonically with
temperature.\cite{wu2,strain}

One can find from the discussion above that the spin dephasing is a
combined effect from the scattering and the inhomogeneous
broadening due to the DP term. The inhomogeneous
broadening induced spin dephasing\cite{wu2,wu3}
 and the counter effect of the
scattering to the inhomogeneous broadening, are both very important and neither can
be neglected. Nevertheless, these effects are either not or
not fully accounted in the simplified
model which is based on the single-particle approximation Eq.\ (\ref{SDT_elec}).
Furthermore, one should notice that the simplified model is based on
the assumption of $|{\bf \Omega}| \tau_p \ll 1$,\cite{Optical} which
is {\em not} always satisfied for holes. To show the
differences between the many-body approach and the earlier treatment,
we also compare our results with those given by the simplified
model which now reads:
\begin{equation}
  \label{SDT_old}
  \frac {1}{\tau_{\lambda}} = \frac{\int^{\infty}_{0}dE_k
    (f_{k\lambda, +}-f_{k\lambda,-})\Gamma_{\lambda}(k)
  }{2\int^{\infty}_{0}dE_k(f_{k\lambda,+}-f_{k\lambda,-})} \ ,
\end{equation}
in which
\begin{eqnarray}
  \label{Gamma}
  \Gamma_{LH}(k)
  &=&k^2[\tau_{1,LH}(\gamma_{53}^{6l6l}k^2+\gamma_{52}^{6l6l}\frac{\pi^2}{a^2})^2
\nonumber\\
&&\mbox{} +
  \tau_{3,LH}(\gamma_{54}^{6l6l})^2]\ ,\\
  \Gamma_{HH}(k) 
  &=&k^6[\tau_{1,HH}(\gamma_{53}^{7h7h})^2 +
  \tau_{3,HH}(\gamma_{54}^{7h7h})^2]  \ ,
\end{eqnarray}
with
\begin{equation}
  \label{tau_p}
  \tau_{l,\lambda}^{-1} = \int^{2\pi}_0
  \sigma_{\lambda}(E_k,\theta)[1-\cos(l\theta)]d\theta\ .
\end{equation}
$\sigma_{\lambda}(E_k,\theta)$ stands for the scattering cross
section of the hole-phonon and the hole-impurity scattering, and
the expressions can be found in Eq.\ (\ref{U_scat}).

In the insets of Figs.\ 2(a) and (b), we plot the corresponding SDTs of LHs and HHs
from the simplified treatments in solid curves.  SDTs from our many-body approach are
plotted in dashed curves. One can see that both the
curvatures and the absolute values are markedly different between the two
treatments. The simplified treatment shows that the SDT of both HHs
and LHs decreases monotonically with temperature
regardless of the impurity densities. Moreover, when the
impurity density increases, the SDT increases very fast. This is because
the single-particle treatment totally ignores the fact that in the
presence of the inhomogeneous broadening, adding a new scattering means
adding a new spin dephasing channel. It also does not treat the counter effect
of the scattering to the inhomogeneous broadening sufficiently. Moreover, it does
not include the Coulomb scattering which we will show in the
next subsection to be very important.
By comparing the SDTs predicted by the two models, one can see
that it is important to study the SDT of holes from the many-body approach.

\subsection{Effect of Coulomb scattering on SDT}
Now we turn to study the effect of the Coulomb scattering in $\dot\rho_{{\bf k}
\lambda,\sigma\sigma^\prime}|_{scatt}$ to the SDT.
It has been shown recently by Wu {\em et al.} that unlike the common belief that the
Coulomb scattering cannot cause spin dephasing, in the presence of inhomogeneous broadening,
it can also lead to spin dephasing\cite{wu2} and
 for electrons in GaAs QWs, the Coulomb scattering is very important
and can markedly increase the SDT.\cite{Weng,MQ} Glazov and Ivchenko have also
drawn the similar conclusion.\cite{mglazov} Since the spin-orbit
coupling of hole system is much larger than that of electron one, it would be
interesting to see how the Coulomb scattering can affect the SDT.
Unlike the case of electron system, here we find that the hole-hole
scattering markedly {\em reduces} the SDT. This is consistent to
the optical dephasing of semiconductors where the Coulomb scattering
gives rise to a stronger optical dephasing.\cite{wu2,Haug}

\begin{figure}
  \centerline{
  \psfig{figure=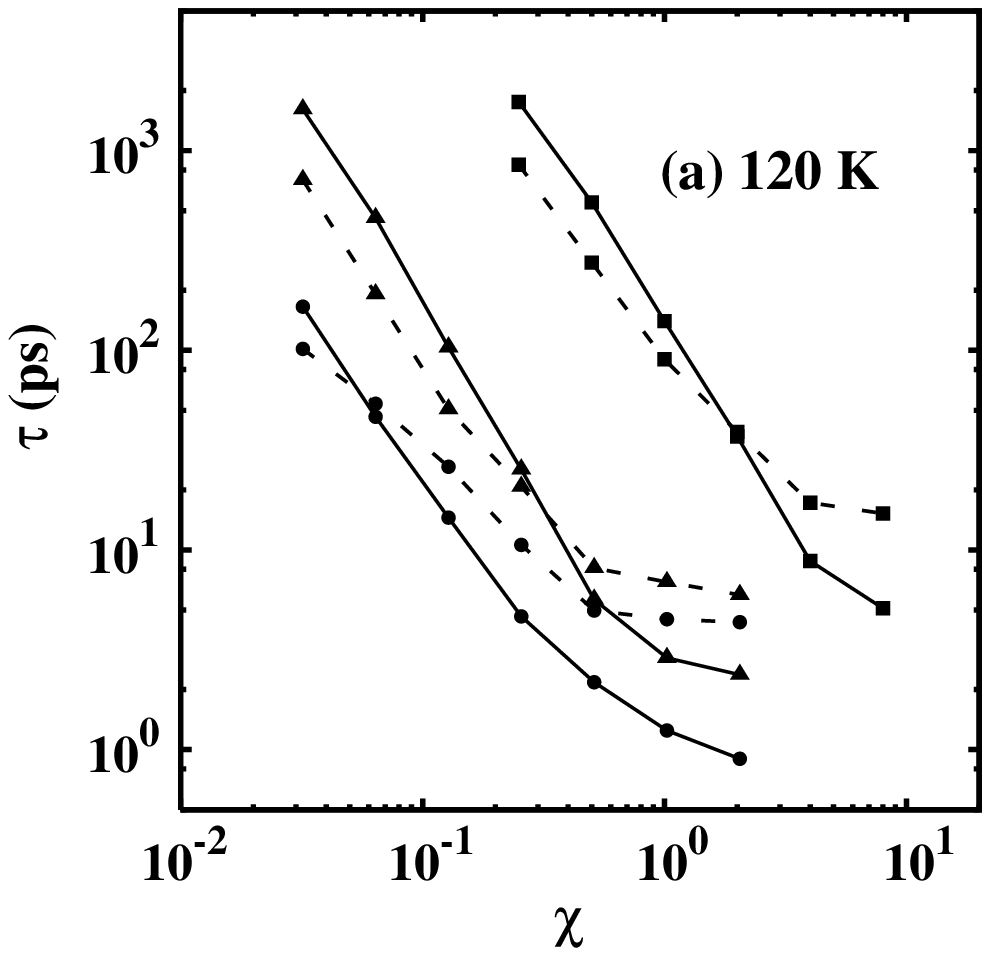,height=5.5cm,angle=0}}
  \vskip 0.5pc
\centerline{\psfig{figure=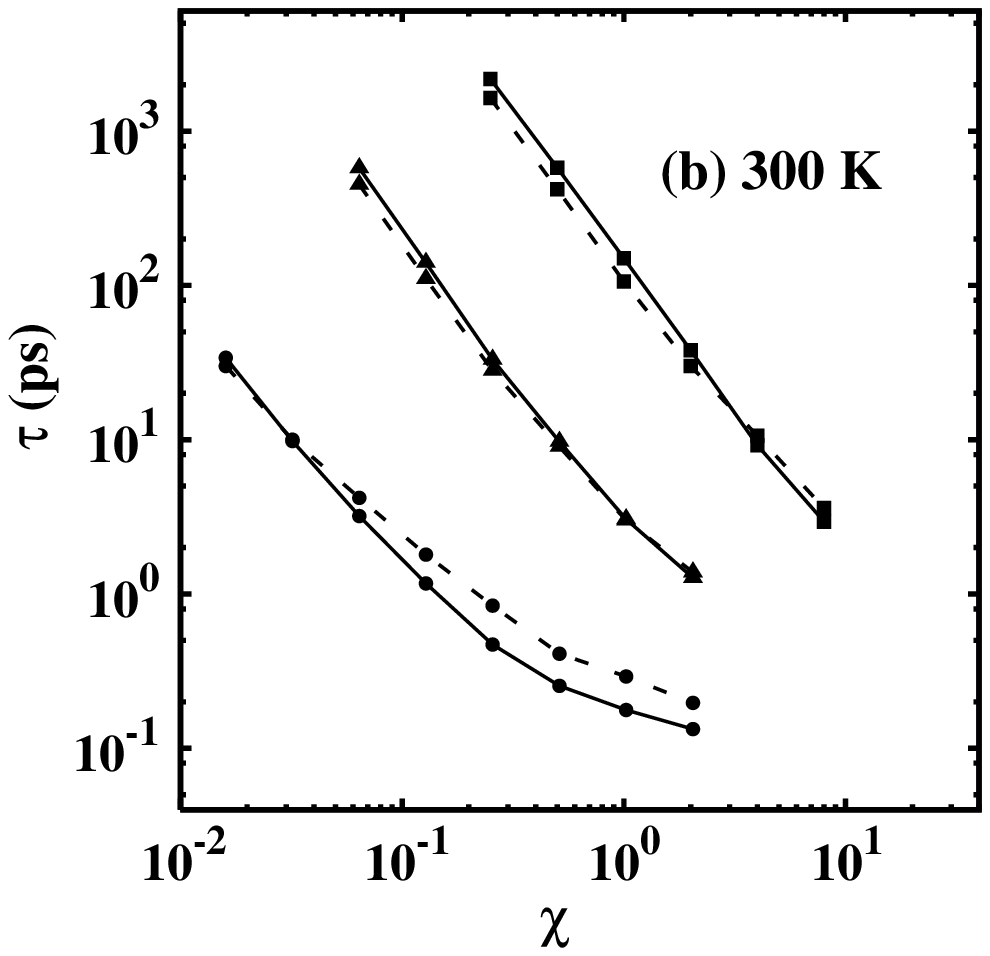,height=5.5cm,angle=0}}
  \caption{SDT  {\em vs.} the scale coefficient of the DP term
$\chi$ for (a) $T=120$\ K and (b) $T=300$\ K. The solid  (dashed)
curves are the results with (without) the Coulomb scattering.
{\tiny $\blacksquare$}: Electrons; {$\blacktriangle$}: LHs;
{$\bullet$}: HHs.}
\end{figure}

In order to understand the difference between the hole system and the previous
electron one, we plot in Fig.\ 3 the SDT of the LH and the HH
as function of a dimensionless scale coefficient of the DP term $\chi$
at $T=120$\ K and 300\ K. Here $\chi$ is introduced by hand in front of the
DP term, {\em i.e.}, $\chi{\bf \Omega}^\lambda({\bf k})$
with $\chi=1$ corresponding to the case of the original DP term. The
solid curves are for the case with both the hole-hole scattering and the
hole-phonon scattering and the dashed ones are for the case with the hole-phonon
scattering only. It is
pointed out here that notwithstanding the fact we sweep $\chi$ through
two orders of magnitude, experimentally the value of the Rashba coefficients can
only  be tuned within a small range by applying an external
electric field,\cite{Rashba_coef1, Rashba_coef2}
and can be determined by analyzing
the Shubnikov-de Haas oscillations.\cite{Wollrab,Luo,Nitta,Heida}
It is seen from the figure that when $\chi=1$, the SDTs of both the HH
and the LH decrease when the Coulomb scattering is included.
However, when one decreases the spin-orbit coupling by decreasing the
scale coefficient $\chi$, one enters a regime where the Coulomb scattering
increases the SDT. This is consistent with our previous observation that
the competing effects of the inhomogeneous broadening and the scattering
in different regimes. In the regime where the inhomogeneous broadening is weak
($|{\bf\Omega}^\lambda|\tau_p \ll 1$), the hole-hole scattering mainly
suppresses the inhomogeneous broadening and consequently raises the SDT.
In the regime where the inhomogeneous broadening is strong,
($|{\bf{\Omega}}^\lambda|\tau_p\gtrapprox 1$),
adding a new scattering provides a new spin
dephasing channel and reduces the SDT. It happens that for hole system, the
spin-orbit coupling\cite{Rashba_coef1} is within the strong
inhomogeneous broadening regime. The same is true also
for the optical dephasing  where adding a new scattering also provides a new
dephasing channel.\cite{wu2,Haug} One also finds
from Fig.\ 3 that the SDT decreases with $\chi$.
This can be easily understood because the
spin dephasing becomes stronger when the spin-orbit coupling is
larger. It is further noted that when the temperature rises,
the effect from the Coulomb scattering
is smaller. This is because the hole-phonon scattering becomes more important
with the increase of temperature.

\begin{figure}
  \centering
  \psfig{figure=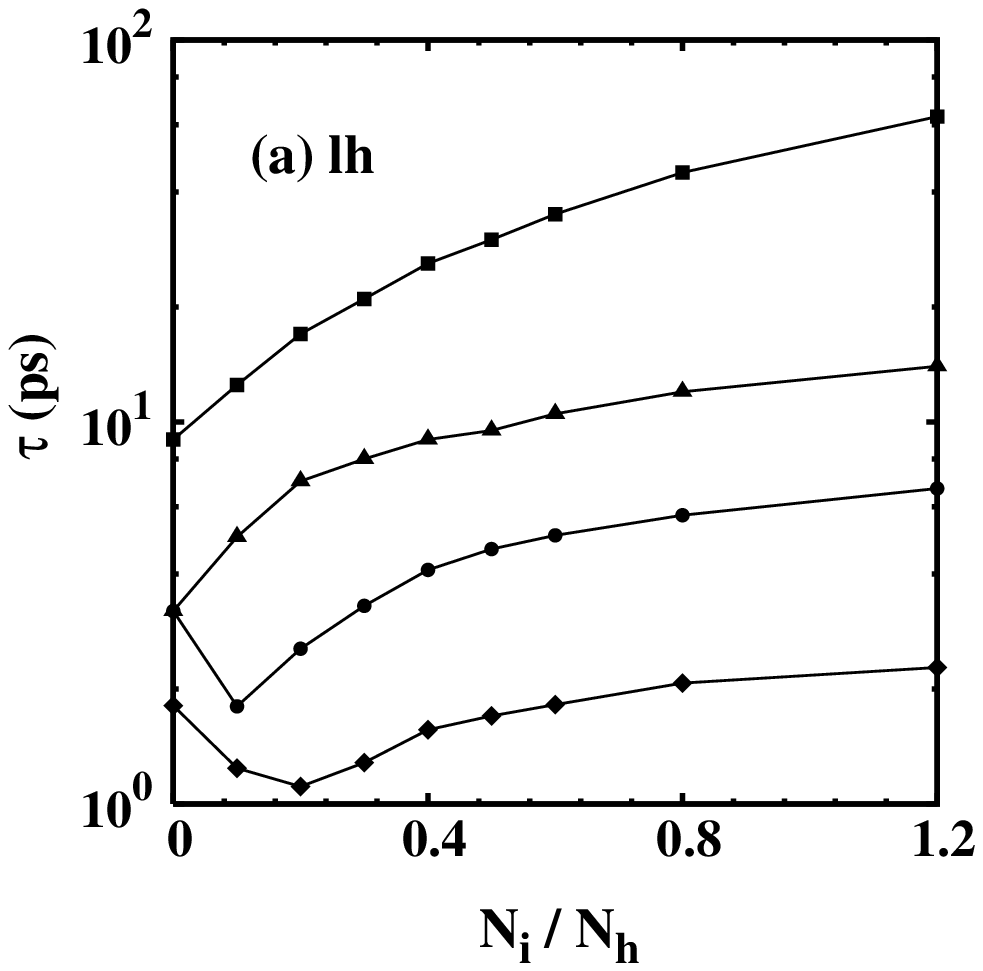,height=5.5cm,angle=0}
  \vskip 0.5pc
  \psfig{figure=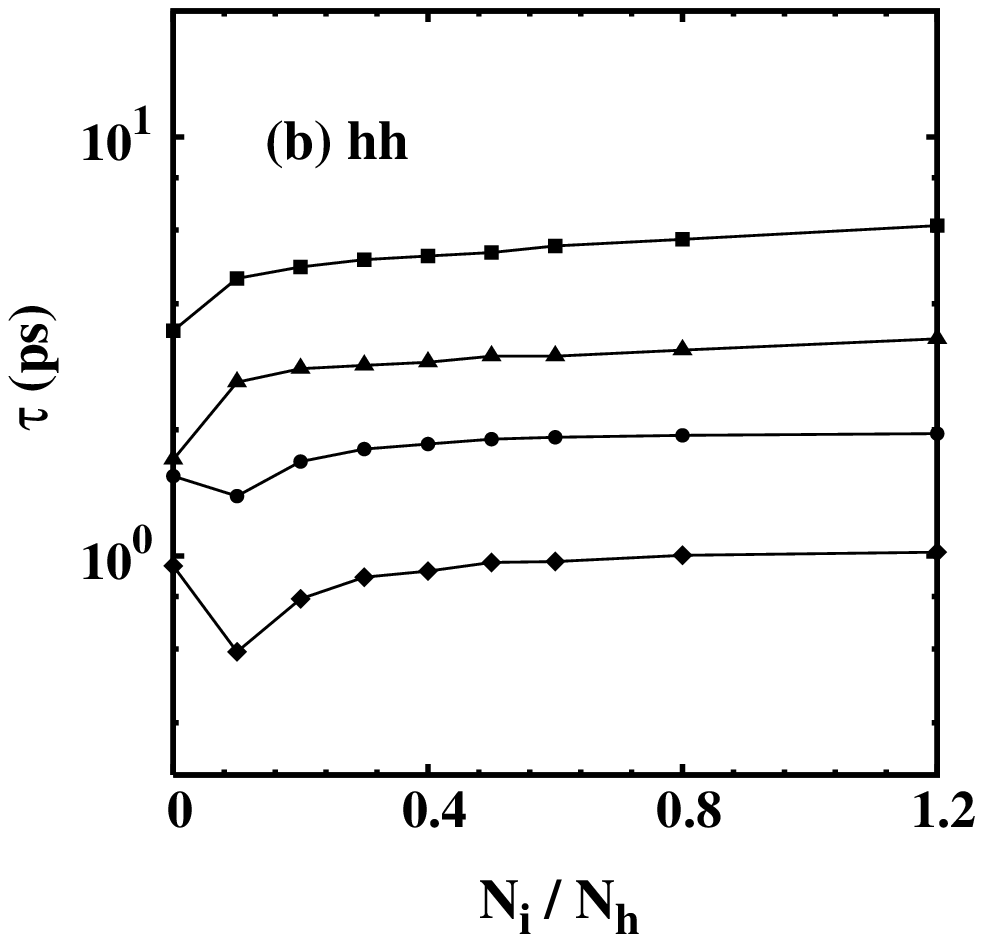,height=5.5cm,angle=0}
  \caption{SDT {\em vs.} the impurity density $N_i/N_h$ with
    different scale coefficients of the DP term $\chi$ at $T= 120$\ K
    for (a) LHs and (b) HHs. {\tiny
      $\blacksquare$}: $ \chi = 0.128$; {$\blacktriangle$}: $ \chi =
    0.256$; {$\bullet$}: $ \chi =  0.512 $; {\small $\blacklozenge$} $
    \chi =  1.024$.}
\end{figure}

In order to compare the hole system with electron one,
 we add a scale coefficient $\chi$ in front of the electron Dresselhaus
term Eq.\ (\ref{elec_DP}) and calculate the SDT of electrons in
a QW ($a=5$\ nm) as function of
$\chi$ with and without the electron-electron Coulomb scattering. The results
are plotted in the same figure for comparison. Similar to the case of holes,
one finds that when the spin-orbit coupling is strong, the Coulomb
scattering reduces the SDT also. It happens that the unscaled DP term ($\chi=1$)
is within the regime of weak inhomogeneous broadening. It is also seen
from the figure that the SDT of holes is much smaller than that of
electrons when $\chi=1$.

\subsection{Impurity density dependence of SDT}
\begin{figure}
  \centering
  \psfig{figure=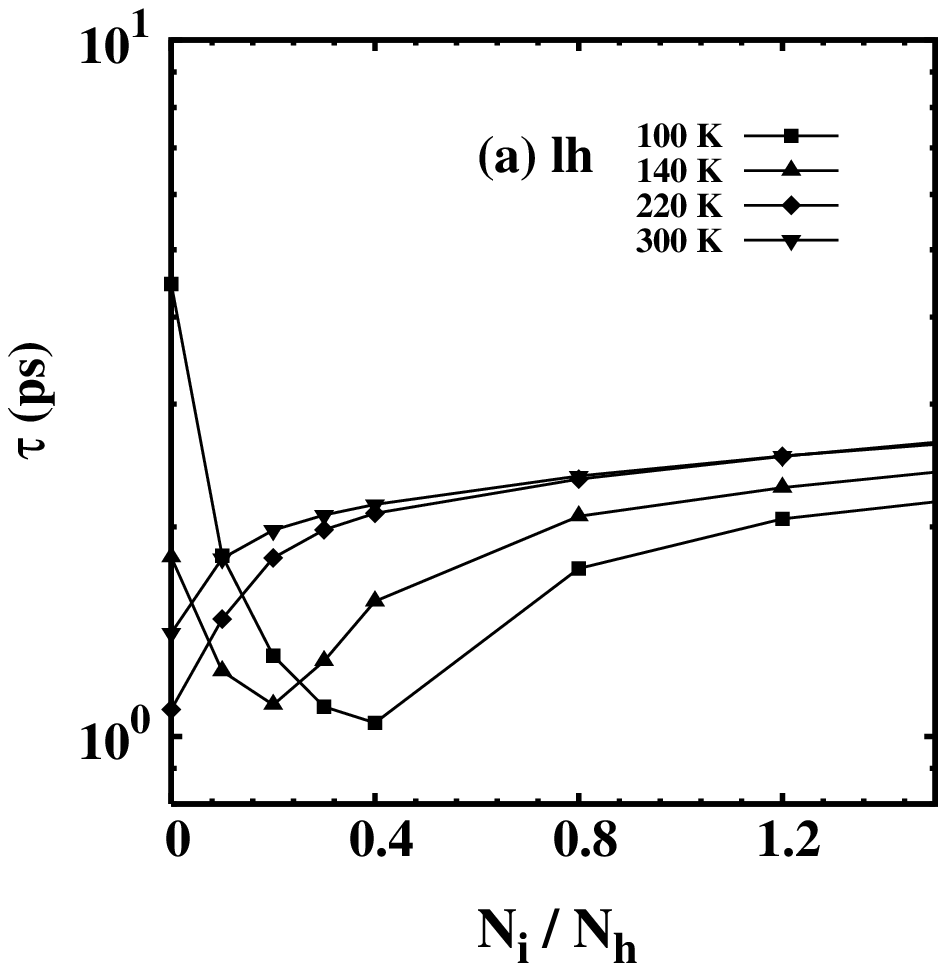,height=5.5cm,angle=0}
  \vskip 0.5pc
  \psfig{figure=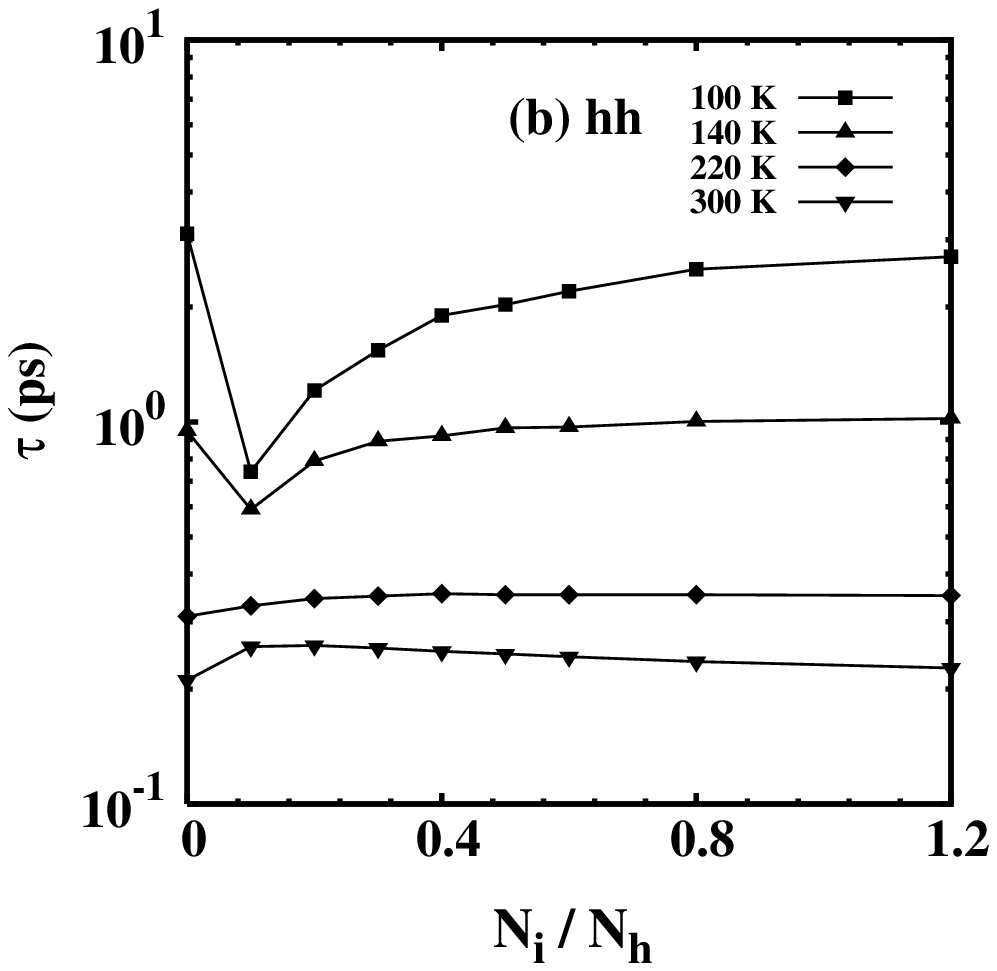,height=5.5cm,angle=0}
  \caption{SDT {\em vs.} the impurity density at
    different temperatures for (a) LHs and (b) HHs. $\chi=1$}
\end{figure}

Now  we turn to study how impurities affect the hole SDT. In Fig.\ 4 we
plot the SDT of LHs as function of the impurity density $N_i$.
It is seen from Fig.\ 4(a) that for LHs when $\chi = 0.128$ or $\chi = 0.256$,
the SDT increases monotonically with the impurity density. Again the spin-orbit
coupling here is in the regime where the inhomogeneous broadening is weak and
the hole-impurity scattering mainly suppresses the
inhomogeneous broadening and raises the SDT.  When the scale coefficient
$\chi = 0.512$ and $N_i = 0$, $|{\bf\Omega}^{LH}({\bf k})| \tau_p({ k})$ is
close to $1$ at the average of $k$: $[|{\bf\Omega}^{LH}({\bf k})|
\tau_p(k)]|_{k=\langle k\rangle}  = 0.32$.  The effect that adding
a new scattering provides a new spin dephasing channel becomes
dominant and the SDT first decreases with the impurity density. This
is similar to the effect of
the Coulomb scattering discussed above while the spin-orbit coupling is
large. However, with the increase of the impurity density, $\tau_p$
gets smaller and $|{\bf\Omega}^{LH}({\bf k})| \tau_p(k)$ gets smaller again.
When $N_i = 0.1N_h$, $[|{\bf\Omega}^{LH}({\bf k})|
\tau_p({k})]|_{k=\langle k\rangle }  = 0.16$,
the SDT reaches a minimum.
Further increasing the impurity density, one enters the
strong scattering regime ($[|{\bf\Omega}^{LH}({\bf
  k})|\tau_p({k})]|_{k=\langle k\rangle }  = 0.049 \ll 1$ at $N_i =
1.0 N_h$), and the SDT keeps increasing with $N_i$ in this regime.
When the scale coefficient $\chi = 1.024$,
$|{\bf\Omega}({\bf k})|$ becomes even larger and the minimum of the SDT
occurs at larger $N_i$. The similar is also true for the case of HHs in
 Fig.\ 4(b). It is noted that in reality $\chi$ is around 1 and the SDT
will first decrease then increase with the impurity density. This is totally
different from the electron case and also different from the prediction of the
single-particle approach where the SDT always increases with the impurity density.

We further investigate the impurity density dependence of the SDT at different
temperatures. Here the scale coefficient $\chi$ is fixed to be 1.
In Fig.\ 5 one finds that when the
temperature is low, the SDT first decreases with $N_i$ as then the
total scattering is weak, and then increases with it after the SDT
reaches a minimum. When the temperature gets higher, the SDT always
increases with $N_i$ as then $|{\bf{\Omega}}^\lambda| \tau_p \ll 1$ can
always be satisfied and the hole-impurity scattering mainly suppresses
the inhomogeneous broadening. 

\subsection{Hole density dependence of SDT}

\begin{figure}
  \centering
  \psfig{figure=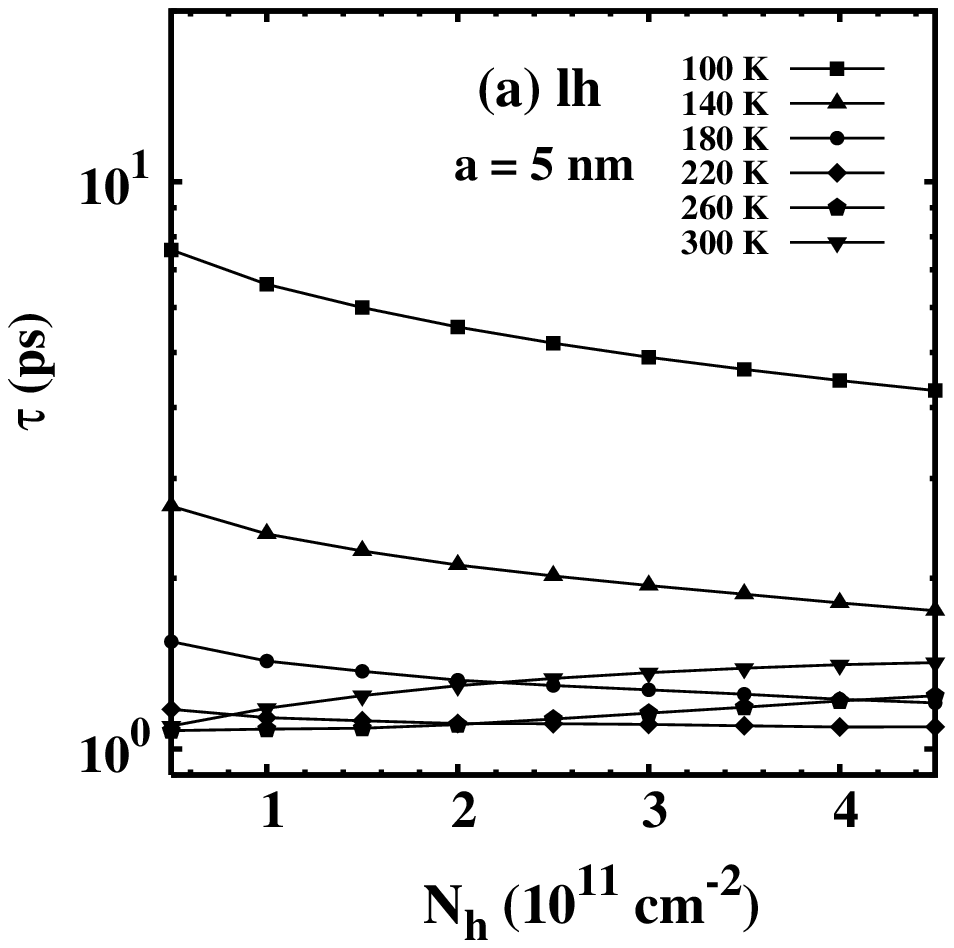,height=5.5cm,angle=0}
  \vskip 0.5pc
  \psfig{figure=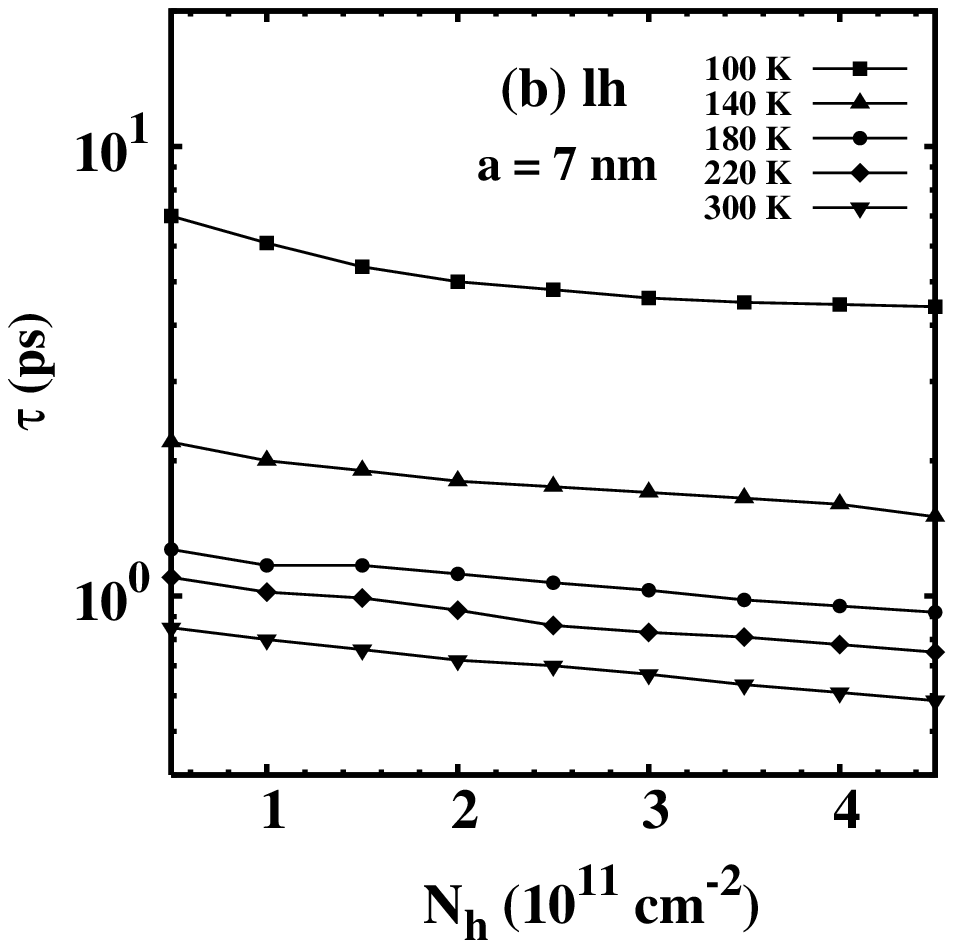,height=5.5cm,angle=0}
  \vskip 0.5pc
  \psfig{figure=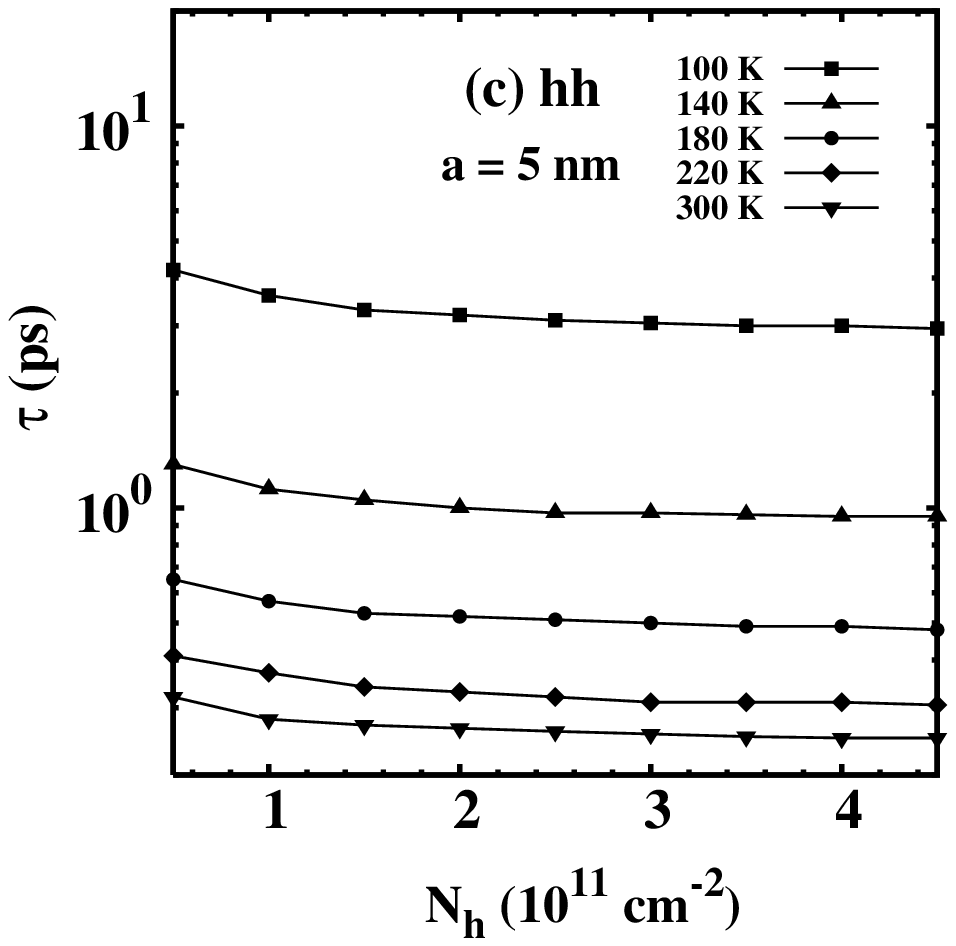,height=5.5cm,angle=0}
  \caption{SDT {\em vs.} the hole density at different  temperatures.
    (a): LHs with $a= 5$\ nm; (b): LHs with
 $a= 7$\ nm; (c): HHs with $a= 5$\ nm.}
\end{figure}

Finally we investigate the hole density dependence of the SDT
at different temperatures and well widthes. Here the
hole-impurity scattering is excluded and $\chi\equiv 1$. In Fig.\
6(a) we plot the SDT of LHs as  function of the LH
density with $a = 5$\ nm. The SDT decreases with the hole
density  when the temperature is low but increases with it when the
temperature is high enough. To understand this result, we first
analyze the Rashba term  [Eqs.\ (\ref{HH_DPx}-\ref{LH_DPz})].
 When $T =
300$\ K, one finds ${\gamma_{52}^{6l6l}E_z}{m_0}\frac{\langle
  k_z^2\rangle}{\langle k_\|^2 \rangle}$ changes from $-2.58$\ nm to
$-1.46$\ nm when $N_h$ changes from $5 \times 10^{10}$\ cm$^{-2}$ to
$4 \times 10^{11}$\ cm$^{-2}$, and the absolute value
becomes even larger when $T = 100$\ K. Therefore,
it can be seen from Table\ \ref{parameter_Rashba} that both
$\gamma_{53}^{6l6l}$ and
$\gamma_{54}^{6l6l}$ are smaller than
$-\gamma_{52}^{6l6l}\frac{\langle
  k_z^2\rangle}{\langle k_\|^2 \rangle}$, and the linear terms
in Eqs.\ (\ref{LH_DPx}) and (\ref{LH_DPy}) are dominant.
Moreover, again $|{\bf\Omega}^{LH}|\tau_p$ is slightly smaller than 1.
Similar to  the case in Sec.\ III B when the linear Rashba term is
dominant, the hole density $N_h$ influences the spin
dephasing through two competing effects: Effect I: The increase of the spin
dephasing due to the increase of the inhomogeneous broadening
with $N_h$ as holes are populated at higher $k$-states at high hole density;
and due to the increase of the scattering  which provides additional
spin dephasing channel. Effect II:
The decrease of the spin dephasing due to the counter effect of the
increased scattering which suppresses the inhomogeneous broadening.
The results shown in Fig.\ 5(a) indicate that when $T \le 220$\ K and the total scattering
is not so strong, Effect I is more important and the SDT decreases
with $N_h$. When the hole density keeps increasing and the total scattering
is further enhanced, Effect II becomes more important and
the SDT increases with $N_h$.

We further plot the SDT of LHs with $a = 7$\ nm in Fig.\ 6(b) where
the cubic terms become more important. For example,
when $T= 300$\ K, ${\gamma_{52}^{6l6l}E_z}{m_0}\frac{\langle
k_z^2\rangle}{\langle k_\|^2 \rangle}$ changes from $ -1.06$\ nm  to
$-0.60$\ nm when $N_h$ changes from $5 \times 10^{10}$\ cm$^{-2}$ to
$4 \times 10^{11}$\ cm$^{-2}$.
One can see from Table\ \ref{parameter_Rashba} that the cubic terms
weighted by $\gamma_{53}^{6l6l}$ are dominant. Similar to the case
in Sec.\ III B, when the cubic Rashba term is dominant,
 the increase of the
inhomogeneous broadening with hole density is much faster than the
counter effect of the scattering and consequently Effect I
always surpasses Effect II with the increase of hole density.
As expected, one finds that the SDT decreases monotonically with
$N_h$. The same is true for HHs in Fig.\ 6(c) where the Rashba term
[Eqs.\ (\ref{HH_DPx}-\ref{HH_DPz})]
includes only the cubic one.

\section{Conclusion}

In conclusion, we have performed a systematic microscopic many-body
investigation on the hole spin dephasing of  $p$-type GaAs
QWs of small well width where the HH and LH bands are well
separated, by constructing a set of kinetic spin Bloch equations based on
the nonequilibrium Green function method. We included the magnetic
field, the Rashba spin-orbit coupling  and all spin conserving scattering
such as the hole-phonon, the hole-nonmagnetic impurity and the hole-hole
scattering. By numerically solving the kinetic equations, we obtained
the time evolution of the distribution functions and the spin coherence of
 holes. The SDT is calculated from the slope of the envelope of the
incoherently summed spin coherence. Differing  from earlier studies
on spin dephasing based on the single-particle approach which only
includes the lowest-order elastic scattering and the anisotropy  from
$-|{\bf\Omega}({\bf k})|$ and $+|{\bf\Omega}({\bf k})|$, this approach
takes full account of the inhomogeneous broadening from different ${\bf k}$-states
of the Rashba term as well as the effect of all the scattering.
Furthermore, this approach is valid
regardless of the strength of scattering whereas the earlier  single-particle
approach is valid
only when the scattering is strong enough, {\em i.e.}, $|{\bf
  \Omega}|\tau_p \ll 1$.
Using this many-body approach, we studied in detail
how the hole spin dephasing are affected by temperature, the hole-hole
Coulomb scattering, the impurity and the hole densities.

We showed that the spin dephasing is mainly affected by two
effects: The inhomogeneous broadening and the scattering. Any effect that
increases the inhomogeneous broadening tends to reduce the SDT.
However, the effect of scattering on the spin dephasing is different when
$|{\bf{\Omega}}|\tau_p \ll 1$ and  $|{\bf{\Omega}}|\tau_p \gtrapprox 1 $:
When $|{\bf{\Omega}}| \tau_p \ll 1$ and therefore the
scattering is strong in comparison to the DP term, the
counter effect of the scattering to the inhomogeneous broadening
is important. In this regime, the scattering
tends to drive carriers to a
more homogeneous state in ${\bf k}$-space and consequently reduces the
inhomogeneous broadening. This tends to increase
the SDT. When $|{\bf{\Omega}}| \tau_p \gtrapprox 1$, the
scattering is weak in comparison to the DP term
(inhomogeneous broadening) and the counter effect can
be neglected, adding a new scattering
provides an additional dephasing channel. In this regime, the
counter effect of the scattering to the inhomogeneous broadening can
be ignored and the scattering  reduces the SDT.
 All the factors, such as temperature, well width, impurity
density and hole density, can affect the inhomogeneous broadening and the scattering and
therefore influence the SDT.

The temperature affects the SDT in two ways: On the
one hand, the
increase of the temperature drives holes to higher ${\bf k}$-states,
and leads to a stronger inhomogeneous broadening. On the other hand,
the scattering is enhanced with the increase of the temperature.
When the linear Rashba term is dominant, such as LHs with
$a = 5$\ nm at the hole density in our investigation, it is shown that
the SDT decreases with $T$ when the temperature is low and the impurity
density is small. This can be understood as it is in the
regime where $|{\bf{\Omega}}^{LH}|$ and $1/ \tau_p$ are comparable and the
increase of the spin dephasing due to
the increase of the inhomogeneous broadening and the increase of the
spin dephasing channel by the increase of the scattering with temperature are
dominant. When the temperature keeps increasing so that the
scattering becomes stronger or when the impurity density is high,
the SDT increases with $T$ when the system enters the regime where $|{\bf{\Omega
 }}^{LH}|\tau_p \ll 1$ and the counter effect of the scattering to
the inhomogeneous broadening becomes
dominant. When the cubic Rashba term is dominant (such as
LHs with $a= 7$\ nm in our investigation) or is the only term (such as HHs),
the SDT decreases monotonically with
temperature as the increase of
the inhomogeneous broadening with temperature is much faster than the
increase of scattering.  These
results are quite different from the case of electrons  where the
spin-orbit coupling is within the regime of weak inhomogeneous broadening
(${\bf{|\Omega |}}\tau_p \ll 1$) and the SDT {\em increases} monotonically
with temperature when the linear DP term is dominant.
We also compared the SDTs predicated by our
many-body approach with the results of the earlier simplified treatment,
and showed that the simplified treatment is inadequate in studying the hole spin
dephasing.

The hole density also influences the inhomogeneous broadening and the
scattering simultaneously. Similar to the case of
temperature dependence, it is shown that for the LHs with $a = 5$\ nm
where the linear Rashba term is dominant,
the SDT decreases with $N_h$ when the temperature is
low because it is in the regime of strong inhomogeneous
broadening, and increases with $N_h$ when the temperature is higher
and the inhomogeneous broadening is weak.
For  LHs with $a = 7$\ nm or HHs where the cubic Rashba
term is the leading/only term, the SDT  decreases monotonically
with $N_h$.

We further showed that the Coulomb scattering
contributes markedly to the SDT. When the inhomogeneous broadening is
stronger than the scattering, the Coulomb scattering enhances the spin
dephasing. Otherwise, it reduces the spin dephasing. In the earlier single-particle
treatment, the Coulomb scattering was considered to be unable to cause spin dephasing.

In the calculation, the magnetic field in the Voigt configuration is taken to be 4\ T.
We found that for hole system, the magnetic field dependence is marginal as the Rashba
term is very large. In this investigation, the Elliott-Yafet mechanism\cite{EY}
 is not included. A full microscopic many-body treatment of the this mechanism
is much more complicated than the DP mechanism and will be published elsewhere.
Up till now there is no  experimental investigation on
the SDT for holes in (001) QWs. Experiments such as spin-echo
measurements\cite{echo} and time-resolved Faraday rotation
measurements\cite{jay} can be used  to measure the SDT.

\begin{acknowledgments}
This work was supported by the Natural Science Foundation of China
under Grant Nos. 90303012 and 10574120, the Natural Science Foundation
of Anhui Province under Grant No. 050460203, the Innovation
Project of Chinese Academy of Sciences and SRFDP. We would like to thank M. Q. Weng
and J. Zhou for their critical reading of this manuscript.
One of the authors (M.W.W.) would like to thank
 Conselho Nacional de Desenvolvimento Cient\'ifico e
Tecnol\'ogico - CNPq, Brazil for financial support and
Ivan Costa da Cunha Lima at Universidade do Estado do Rio de Janeiro (Brazil)
for hospitality where this work was finalized.
\end{acknowledgments}

\begin{appendix}

\section{Effect of Scattering to SDT}

In Sec.\ III B we pointed out that when the
scattering is weak in comparison to the DP effective
field (inhomogeneous broadening), the main effect of scattering is
to add an additional dephasing channel. Therefore the scattering here reduces
the SDT. To demonstrate this effect, we now study the spin dephasing in the
limiting case of no scattering included in the calculation. In Fig.\ 7
we show the temporal evolution of the incoherently summed spin coherence
$\rho_{LH}= {\sum_{\bf k}|\rho_{{\bf k}LH}(t)}|$ for LHs with
total LH density $N_h = 4
\times 10^{11}$\ cm$^{-2}$ and $T = 300$\ K. The coherently summed
spin coherence $\rho^\prime_{LH} = |{\sum_{\bf k}\rho_{{\bf
      k}LH}(t)}|$  is also plotted for comparison. One can see
that the amplitude of $\rho^\prime_{LH}$  oscillates  and decays to
zero very quickly due to the interference caused by the momentum
dependence of the DP term. However, the incoherently summed
spin coherence $\rho_{LH}$ does not decay, which means that there
is no irreversible dephasing.\cite{Haug,wu1,Kuhn} This is consistent to the
fact of no scattering as there is no dissipation process here.
By adding a  scattering, one introduces the dissipation into the
system which causes an irreversible dephasing. This can be
seen in the same figure where  the incoherently summed spin coherence $\rho_{LH}$
is plotted with the impurity density $N_i=0.01N_h$.
One finds that the $\rho_{LH}$ now decays with time, although much slower
than the one in Fig.\ 1 where all the scattering is included.

\section{A Simplified Analytical Analysis of SDT}

In Sec.\ III B we pointed out that the
ratio of the DP term  to the
scattering rate determines the way how the scattering affects the spin
dephasing. To reveal part of this effect {\em analytically},
we now study a much simplified case
with only the hole-impurity scattering and the
 dominant part of the DP term, {\em i.e.}, for LHs
we include only the linear Rashba term weighted by $\gamma_{52}^{6l6l}$ and
for HHs only the cubic Rashba term weighted by $\gamma_{54}^{7h7h}$. Furthermore,
we will also neglect the inhomogeneous broadening later.

\begin{figure}
  \centerline{  \psfig{figure=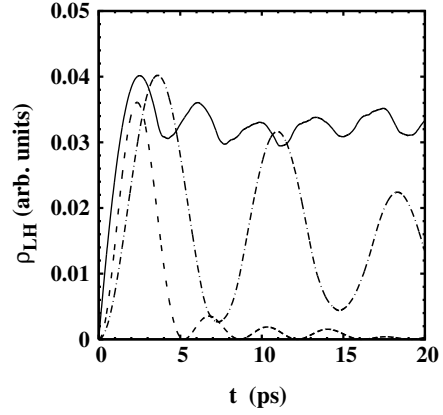,height=5.5cm,angle=0}}
  \caption{Temporal  evolution of the incoherently summed spin
  coherence $\rho_{LH}(t)$ (solid curve) and the
  coherently summed spin coherence $\rho^\prime_{LH}(t)$ (Dotted curve) without
any scattering and the temporal evolution of the incoherently summed spin
  coherence $\rho_{LH}(t)$ with only hole-impurity scattering with $N_i=0.01N_h$
(Chained Curve).
}
\end{figure}

First we consider the HH case. We expand $2\times2$ density
matrix $\rho_{{\bf k} HH}$ as follows:
\begin{equation}
  \label{expand_rho}
  \rho_{{\bf k} HH} = \sum\limits_l \rho_{HH,l}(k) A_l(\theta_{\bf k})
\end{equation}
with $A_l(\theta_{\bf k}) = \frac{1}{\sqrt{2\pi}} e^{il\theta_{\bf k}}$.
The coherent terms of the kinetic spin Bloch equations [Eqs.\ (\ref{coh_f})
and (\ref{coh_rho})] can be written into the matrix commutator
as $\dot{\bf \rho}_{{\bf k},HH}|_{coh} = i[H^{HH}_{s}({\bf k}),\rho_{{\bf k},HH}]$
      with $H^{HH}_{s}({\bf k}) = \frac{1}{2}
\mbox{\boldmath$\sigma$\unboldmath} \cdot {\bf
\Omega}^{HH}({\bf k})$. It is noted that here we neglected the Coulomb
Hartree-Fock term. Furthermore, we expand $H^{HH}_{s}$ as
$H^{HH}_{s}({\bf k}) = \sum_l
      H^{HH}_{s,l}(k) A_l(\theta_{\bf k})$. Therefore the coherent
term in the matrix form can be written as:
\begin{equation}
  \label{coh_1}
  \dot{\bf \rho}_{{\bf k},HH}|_{coh} =
\sum_{l,l_1}\frac{i}{\sqrt{2\pi}}
[H^{HH}_{s,l-l_1}({\bf k}),
\rho_{HH,l_1}(k)]A_l(\theta_{\bf k}) \ .
\end{equation}
With only the dominant part of the DP term
(term weighted by $\gamma_{54}^{7h7h}$) included,
$H^{HH}_s({\bf k})$ is expanded as:
\begin{eqnarray}
  \label{expand_H3}
&&  H_{s,3}^{HH}= i S^{\dagger}\gamma_{54}^{7h7h} { E}_z k_{\|}^3\ ,\\
\label{expand_H-3}
&&  H_{s,-3}^{HH}= - i S\gamma_{54}^{7h7h} { E}_z k_{\|}^3\ ,\\
\label{expand_Hl}
&&  H_{s,l\ne \pm 3}^{HH} = 0\ ,
\end{eqnarray}
in which $S = \frac{1}{2}(\sigma_x -i\sigma_y)$. Substituting Eqs.\
(\ref{expand_H3}-\ref{expand_Hl}) into Eq.\ (\ref{coh_1}), one obtains
\begin{eqnarray}
  \label{coh_2}
  \dot{\bf \rho}_{{\bf k} HH}|_{coh}& =& \frac{1}{\sqrt{2\pi}}
\gamma_{54}^{7h7h} { E}_z k_{\|}^3
\sum_l \Big([S, \rho_{HH, l+3}(k)] \nonumber \\
&&\mbox{} - [S^{\dagger}, \rho_{HH, l-3}(k)]\Big)A_l(\theta_{\bf k}) \ .
\end{eqnarray}

Similarly one can expand the scattering term as:
\begin{equation}
  \label{scat_1}
  \dot{\bf \rho}_{{\bf k} HH}|_{scat} = \sum\limits_l \rho_{HH,l}(k) U_l^2(k)
  A_l(\theta_{\bf}) \ ,
\end{equation}
for the elastic scattering with
\begin{equation}
  \label{U_scat}
  U_l^2(k) = 2\pi N_i\frac{m^\ast}{\hbar^2}
  \int_0^{2\pi}\frac{d\theta}{(2\pi)^2}
  U_{q(\theta)}^2(1-\cos{l\theta}) \ .
\end{equation}
Here $q(\theta) = \sqrt{2k^2(1-\cos{\theta})}$, and $U_{q(\theta)}^2
= \sum_{q_z} \{4\pi e^2/[\kappa_0 (q^2(\theta)+ q_z^2)]\}^2
|I(iq_z)|^2 $ is the hole-impurity
scattering matrix element and $|I(iq_z)|^2 = \pi^2
\sin^2 y/[y^2(y^2-\pi^2)^2]$ with $y = q_z a/2$ is the form
factor.

Now we can expand the spin Bloch equations Eq.\ (\ref{Bloch_eq})
into $\rho_{HH,l}(k)$ as follows:
\begin{eqnarray}
  \label{Bloch_eq2}
&&\dot{\rho}_{HH,l}(k) + \gamma_{54}^{7h7h} { E}_z k_{\|}^3
  \frac{1}{\sqrt{2\pi}} \Big([S, \rho_{HH,l+3}(k)]
 \nonumber \\
&&\mbox{}\hspace{0.6cm} - [S^{\dagger},\rho_{HH,l-3}(k)]\Big)
= -  U_l^2(k)\rho_{HH,l}(k) \ .
\end{eqnarray}
In order to find the solution, we multiply $\mbox{\boldmath$\sigma$\unboldmath}
= (\sigma_x,\sigma_y, \sigma_z)$ to both sides of this
equation and calculate the trace. By defining the HH ``spin'' vector
to be  $\mbox{Tr}(\rho_{HH,l}(k)
\mbox{\boldmath$\sigma$\unboldmath})\equiv
{{\bf S}_{HH,l}(k)}$, one can rewrite Eq.\ (\ref{Bloch_eq2}) into
\begin{eqnarray}
  \label{Bloch_eq3}
&&\dot{\bf S}_{HH,l}(k) + \gamma_{54}^{7h7h} { E}_z k_{\|}^3
  \frac{1}{\sqrt{2\pi}} \Big({\cal F}\cdot{\bf S}_{HH,l+3}(k)\nonumber \\
&&\mbox{}\hspace{0.3cm}
-{\cal F}^{\dagger}\cdot{\bf S}_{HH, l-3}(k)\Big) = -  U_l^2(k){\bf S}_{HH,l}(k) \ ,
\end{eqnarray}
thanks to the relation
$\mbox{Tr}([S, \rho_{HH, l-1}(k)]\mbox{\boldmath$\sigma$\unboldmath})
= \mbox{Tr}(\rho_{HH, l-1}(k)[\mbox{\boldmath$\sigma$\unboldmath}, S])$.
In Eq.\ (\ref{Bloch_eq3}) the tensor ${\cal F}$ reads
\begin{equation}
  \label{f_matrix}
  {\cal F} = \left( \begin{array}{ccc}  0 & 0 & 1 \\ 0 & 0 &
  -i \\  -1 & i & 0
  \end{array} \right)\ .
\end{equation}
One finds that ${{\bf S}_{HH,l}}(k)$ is only related to ${{\bf S}_{HH, l\pm 3}}(k)$.
By considering the lowest orders with $l = 0$, and $\pm 3$ and defining
${\bf\cal S}_{HH}(k) = ({\bf S}_{HH, -3}(k),{\bf S}_{HH,0}(k),{\bf S}_{HH, 3}(k))^{T}$,
Eq.\ (\ref{Bloch_eq3}) can be written as:
\begin{equation}
  \label{Bloch_eq_final}
  \dot{\bf \cal S}_{HH}(k) + (\frac{\gamma_{54}^{7h7h} { E}_z k_{\|}^3}
  {\sqrt{2\pi}} {\cal G} + {\cal U} ){\bf\cal S}_{HH}(k) =0 \ ,
\end{equation}
in which
\begin{eqnarray}
  \label{F}
 {\cal G}& =& \left( \begin{array}{ccc}  0 &{\cal F}& 0 \\ -{\cal F}^{\dagger} & 0 &
 {\cal F}\\  0 & -{\cal F}^{\dagger} & 0
  \end{array} \right)\ ,\\
  \label{U_matrix}
{\cal  U} &=& U_3^2(k) \left( \begin{array}{ccc}  1 & 0 & 0 \\ 0 & 0 &
  0 \\  0 & 0 & 1
  \end{array} \right)\ .
\end{eqnarray}
In Eq.\ (\ref{U_matrix}) we have used the relations $U_3^2(k) = U_{-3}^2(k)$ and
$U_0^2(k) = 0$.

Now one can solve Eq.\ (\ref{Bloch_eq_final}) analytically. To
reveal the main characteristic analytically, we make the assumption
that the spin relaxation/dephasing occurs mainly around the Fermi surface
and $\theta_{{\bf k}_f} = 0$. By doing so, one throws away the interference
between different ${\bf k}$-states [except for the
states with $\theta_{{\bf k}_f}$($=0$ here),
and $\theta_{{\bf k}_f}\pm2\pi/3$], and therefore the inhomogeneous
broadening. Then
the HH spin ${\bf S}_{HH} \equiv\sum\limits_{l=-3,0,3}
(S_{HH,l}^x,S_{HH,l}^y,S_{HH,l}^z)$ has the form:
\begin{eqnarray}
  \label{Sx}
&&S_{HH}^x= 0 \ , \\
\label{Sy}
&&S_{HH}^y=
\frac{2e^{-\frac{1}{2}t_1(x+\sqrt{x^2-16})}}{\sqrt{x^2-16}}(e^{t_1\sqrt{x^2-16}}
-1 )S_{0} \ , \\
\label{Sz}
&&S_{HH}^z=
  \frac{e^{-\frac{1}{2}t_1(x+\sqrt{x^2-16})}}{2\sqrt{x^2-16}}
  [x(e^{t_1\sqrt{x^2-16}} -1 ) \nonumber \\
&&\mbox{}\hspace{0.9cm}+ \sqrt{x^2 -16}
  (e^{t_1\sqrt{x^2-16}} + 1 )]S_{0}\ ,
\end{eqnarray}
 in which $x = \frac{U_3^2(k_f) \sqrt{2\pi}}{\gamma_{54}^{7h7h} {
     E}_z k_f^3}$ is proportional to the ratio of the
scattering rate to the DP term, $t_1 = \frac{\gamma_{54}^{7h7h} { E}_z
   k_f^3}{\sqrt{2\pi}} t $ and $S_{0}$ represents the initial spin polarization
along the $z$-axis. One can see from Eq.\ (\ref{Sy}) that when
 $x<4$, the SDT is proportional to $1/x$  and {\em decreases} with $x$ whereas when
 $x > 4$, the SDT is proportional to $1/(x-\sqrt{x^2-16})$ and {\em increases}
 with $x$. This result indicates that the scattering reduces the
 SDT when the scattering is weak in comparison to the DP effective
 field but increases the SDT when the scattering is strong in comparison
 to the DP effective field. Moreover, when $U_3=0$, {\em i.e.}, there is no
 scattering, $x=0$ and consequently there is no spin dephasing. This is
 consistent with the numerical result presented in Appendix A.

 Similarly one can derive the equation for spin of LHs ${\bf S}_{LH}$
 with only the linear part of the Rashba term included. One gets the
 same equations Eqs.\ (\ref{Sx}-\ref{Sz}) but with $x = \frac{U_3^2(k_f)
   \sqrt{2\pi}}{\gamma_{52}^{6l6l} { E}_z \langle k_z^2\rangle k_f}$ and $t_1 =
 \frac{\gamma_{52}^{6l6l} { E}_z  \langle k_z^2\rangle k_f}{\sqrt{2\pi}} t $.

These results coincide qualitatively with the results shown  in
 Fig.\ 4: The SDT first decreases then increases with the
hole-impurity scattering when the spin-orbit coupling is
strong; but increases monotonically with the scattering when the spin-orbit coupling
is weak. Furthermore, by making the approximation that the hole-phonon
scattering is also an elastic scattering and by including the hole-phonon
scattering in the scattering term, $U_q^2$ in Eq.\
({\ref{U_scat}}) can be modified as:
\begin{eqnarray}
  \label{g_U_scat}
  U_l^2(k) &=&2\pi\frac{m^\ast}{\hbar^2}
  \int_0^{2\pi}\frac{d\theta}{(2\pi)^2} \
[N_i  U_q^2  + (1+2N_q)g_q^2]\nonumber \\ &\times& (1-\cos{l\theta}) \ .
\end{eqnarray}
Here $g_q^2 = \sum\limits_{q_z}\{2\pi e^2 \Omega_{\mbox{\tiny LO}}/[
(q^2+ q_z^2)]\}(1/\kappa_{\infty}-1/\kappa_0) |I(iq_z)|^2 $ is the hole-phonon
interaction matrix element, and $N_q = 1/[\mbox{exp}( \Omega_{\mbox{\tiny
    LO}} /k_BT) - 1]$ is the Bose distribution of the LO phonon.
Then the results also coincide qualitatively with those in Fig.\
5(b): the SDT  first decreases then increases with the
hole-impurity scattering when the total scattering is weak, but
always increases with scattering when the total scattering is strong. Finally we
point out that as we do not include the inhomogeneous broadening and all the scattering
in this simplified model, many other features predicted in the text cannot be
obtained here.

\end{appendix}


\begin{thebibliography}{0}
\bibitem{comment} As shown in Ref.\ \onlinecite{wu3}, if the DP
term is isotropic as $|{\bf \Omega}({\bf k})|$,
then the spin dephasing predicted by the lowest order as
Eq.\ (\ref{SDT_elec}) is zero.
\end{thebibliography}
\end{document}